\documentclass[aoas,MSNbibl,nameyear,seceqn,dvips]{arximspdf}
\usepackage{graphicx}
\doi{10.1214/12-AOAS609} 
\volume{7}
\issue{2}
\pubyear{2013}
\firstpage{904}
\lastpage{935}

\makeatletter

\newproclaim{Remarks}{Remarks}

\newcommand{\cal}{\mathcal}

\newcommand{\ve}[1]{\bolds{#1}}

\makeatother

\begin{document}
\begin{frontmatter}

\title{Multiscale adaptive smoothing models for the
hemodynamic response function in fMRI}
\runtitle{Multiscale adaptive smoothing models}

\begin{aug}
\author[A]{\fnms{Jiaping} \snm{Wang}\ead[label=e1]{jwang@bios.unc.edu}},
\author[A]{\fnms{Hongtu} \snm{Zhu}\corref{}\thanksref{t1}\ead[label=e2]{hzhu@bios.unc.edu}},
\author[B]{\fnms{Jianqing} \snm{Fan}\thanksref{t3}\ead[label=e3]{jqfan@princeton.edu}},
\author[C]{\fnms{Kelly}~\snm{Giovanello}\ead[label=e4]{kgio@email.unc.edu}}
\and
\author[D]{\fnms{Weili} \snm{Lin}\thanksref{t2}\ead[label=e5]{weili\_lin@med.unc.edu}}
\runauthor{J. Wang et al.}
\affiliation{University of North Carolina at Chapel Hill,
University of North Carolina at Chapel Hill,
Princeton University,
University of North Carolina at Chapel Hill and
University of North Carolina at Chapel Hill}
\address[A]{J. Wang\\
H. Zhu \\
Department of Biostatistics\\
University of North Carolina\\
\quad at Chapel Hill\\
Chapel Hill, North Carolina 27599\\
USA\\
\printead{e1}\\
\hphantom{E-mail: }\printead*{e2}}
\address[B]{J. Fan\\
Department of Operations Research\hspace*{30pt}\\
\quad and Financial Engineering \\
Princeton University\\
Princeton, New Jersey 08540\\
USA\\
\printead{e3}}
\address[C]{K. Giovanello\\
Department of Psychology \\
University of North Carolina\\
\quad at Chapel Hill\\
Chapel Hill, North Carolina 27599\\
USA\\
\printead{e4}}
\address[D]{W. Lin\\
Department of Radiology\\
\quad and Biomedical Research Imaging Center\\
University of North Carolina\\
\quad at Chapel Hill\\
Chapel Hill, North Carolina 27599\\
USA\\
\printead{e5}} 
\end{aug}

\thankstext{t1}{Supported in part by NIH Grants RR025747-01,
P01CA142538-01, EB005149-01 and MH086633.}

\thankstext{t3}{Supported in part NIH Grant R01-GM07261, NSF
Grants DMS-07-04337 and DMS-07-14554.}

\thankstext{t2}{Supported in part NIH Grant R01NS055754.}

\received{\smonth{12} \syear{2011}}
\revised{\smonth{10} \syear{2012}}

%
\begin{abstract}
In the event-related functional magnetic resonance imaging (fMRI) data
analysis, there is an extensive interest in accurately and robustly
estimating the hemodynamic response function (HRF) and its associated
statistics (e.g., the magnitude and duration of the activation). Most
methods to date are developed in the time domain and they have utilized
almost exclusively the temporal information of fMRI data without
accounting for the spatial information. The aim of this paper is to
develop a multiscale adaptive smoothing model (MASM) in the frequency
domain by integrating the spatial and frequency information to
adaptively and accurately estimate HRFs pertaining to each stimulus
sequence across all voxels in a three-dimensional (3D) volume. We use
two sets of simulation studies and a real data set to examine the
finite sample performance of MASM in estimating HRFs. Our real and
simulated data analyses confirm that MASM outperforms several other
state-of-the-art methods, such as the smooth finite impulse response
(sFIR) model.
\end{abstract}

%

\end{frontmatter}

\section{Introduction}\label{sec1}

Since the early 1990s, functional magnetic resonance imaging (fMRI) has
been extensively used in the brain mapping field because of its
relatively low invasiveness, absence of radiation exposure, relatively
wide availability, relatively high spatial and temporal resolution,
and, importantly, signal fidelity. It has become the tool of choice in
behavioral and cognitive neuroscience for understanding functional
segregation and integration of different brain regions in a single
subject and across different populations [\citet{FristonEtal2009},
\citet{Friston2007}, \citet{HuettelEtal2004}]. It commonly uses
blood oxygenation level-dependent (BOLD) contrast
[\citet{OgawaEtal1992}] to measure the hemodynamic response (e.g.,
change in blood oxygenation level) related to neural activity in the
brain or spinal cord of humans or animals. Thus, most fMRI researches
correlate the BOLD signal elicited by some specific cognitive process
with the underlying unobserved neuronal activation.

%
\begin{figure}

\includegraphics{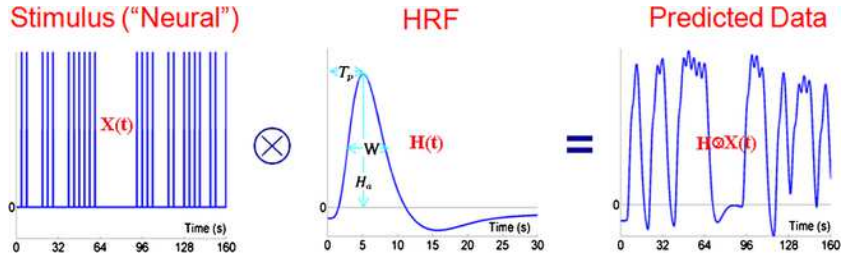}

\caption{A diagram of the fMRI signals generated by the circular
convolution between
the stimulus sequence $X(t)$ and the hidden HRF $H(t)$ without
specifying the voxel ${\mathbf d}$ for notational simplicity. In the
diagram of
$H(t)$, $H_a$ is the response amplitude/height, $T_p$ is the
time-to-peak, and $W$ is the full-width at half-max.}
\label{figure1}
\end{figure}

In the modeling literature of fMRI data, a linear time invariant (LTI)
system is commonly implemented to model the linear relationship between
a stimulus sequence and the BOLD signal [\citet{BoyntonEtal1996},
\citet{FristonEtal1994}]. Specifically, the BOLD signal at time
$t$ and voxel ${\mathbf d}$, denoted as $Y(t, {\mathbf d})$, is the
convolution of a stimulus function, denoted as $X(t)$, and a
hemodynamic response function (HRF), denoted as $H(t, {\mathbf d})$,
plus an error process, denoted as $\varepsilon(t, {\mathbf d})$. See
Figure \ref{figure1} for an illustration of LTI. While nonlinearities
in the BOLD signal are predominant for stimuli with short separations
[\citet{BoyntonEtal1996}, \citet{BuxtonEtal1998}], it has
been shown that LTI is a reasonable assumption in a wide range of
situations [\citet{Glover1999}, \citet{FristonEtal1994}].
Furthermore, with the advent of event-related fMRI, it is possible to
estimate the shape of HRF elicited by cognitive events. Given the shape
of the estimated HRF, it is also important to extract several HRF
measures of psychological interest, including the response
amplitude/height ($H_a$), time-to-peak ($T_p$) and full-width at
half-max ($W$) in HRF (see the definitions of $H_a$, $T_p$ and $W$ in
Figure \ref{figure1}), which may be correlated with the intensity,
onset latency and duration of the underlying brain metabolic activity
under various experimental manipulations
[\citet{BellgowanEtal2003}, \citet{FormisanoAndGoebel2003},
\citet{RichterEtal2000}, \citet{LindquistAndWager2007}]. It has been
shown that minor amounts of mis-modeled HRFs or BOLD signals can lead
to severe loss in power and validity [\citet{LindquistAndWager2007},
\citet{LohEtal2008}, \citet{CasanovaEtal2008},
\citet{LindquistEtal2009}]. Thus, it is important to obtain an
accurate estimate of the HRF shape, which is the focus of this paper.

In the last decade, dozens of time domain HRF models have been proposed
and implemented in the existing neuroimaging software platforms,
including statistical parametric mapping (SPM)
(\href{http://www.fil.ion.ucl.ac.uk/spm/}{www.fil.ion.ucl.ac.uk/spm/})
and the FMRIB Software Library (FSL)
(\href{http://www.fmrib.ox.ac.uk/fsl/}{www.fmrib.ox.ac.uk/fsl/}), among
many others. For instance, SPM uses a combination of the canonical HRF
and its derivatives with respect to time and dispersion
[\citet{FristonEtal1994}, \citet{HensonEtal2002}]. Other
approaches include a finite impulse response (FIR) basis set
[\citet{Glover1999}, \citet{OllingerEtal2001}], the use of
basis sets composed of principal components
[\citet{AguirreEtal1998}, \citet{WoolrichEtal2004}], spline
basis sets [\citet{ZhangEtal2007}], a canonical function with free
parameters for magnitude and onset/peak delay
[\citet{LindquistAndWager2007}, \citet{MiezinEtal2000}], the
Bayesian method [\citet{Genovese2000}, \citet{r23},
\citet{KimEtal2010}] and several regularization-based techniques
[\citet{VakarinEtal2007}, \citet{CasanovaEtal2008}].
Particularly, \citet{CasanovaEtal2008} have shown that the
estimates of HRF can be sensitive to the temporal correlation
assumption of the error process.

Only few HRF models are studied in the frequency domain. The basic
idea of these frequency domain models is to transform the original
fMRI signal into the frequency coefficients and then develop a
statistical model based on these coefficients. For instance,
\citet{LangeAndZeger1997} developed a model in the frequency domain
along with a two-parameter gamma function as the HRF model. For
experimental designs with periodic stimuli,
\citet{MarchiniAndRipley2000} proposed a model in the frequency
domain with a fixed HRF. Recently, \citet{BaiEtal2009} used a
nonparametric method to estimate HRF based on point processes
[\citet{Brillinger1974}]. In comparison to the time domain approaches,
these frequency domain models are less sensitive to the temporal
correlation assumption of the error process
[\citet{MarchiniAndRipley2000}], since these Fourier coefficients are
approximately uncorrelated across frequencies.

Almost all of the HRF models discussed above have exclusively
estimated HRF on a voxel-wise basis and ignored the fact that fMRI
data are spatially dependent in nature. Specifically, as is often
the case in many fMRI studies, we observe spatially contiguous
effect regions with rather sharp edges. There have been several
attempts to address the issue of spatial dependence in fMRI. One
approach is to apply a smoothing step before individually estimating
HRF in each voxel of the 3D volume. As pointed out by
\citet{YueEtal2010} and \citet{LiEtal2011}, most smoothing
methods, however, are
independent of the imaging data and apply the same amount of
smoothness throughout the whole image. These smoothing methods can
blur the information near the edges of the effect regions and thus
dramatically increase the number of false positives and false
negatives.
An alternative approach is to explicitly model spatial dependence
among spatially connected voxels by using conditional autoregressive
(CAR) and Markov random field (MRF), among others
[\citet{Besag1986}, \citet{Bowman2007}]. However, besides a
specific type of
correlation structure, such as MRF, calculating the normalizing
factor of MRF and estimating spatial correlation for a large number
of voxels in the 3D volume are computationally intensive.

The goal of this paper is to develop a multiscale adaptive smoothing
model (MASM) in the frequency domain to adaptively construct an
accurate nonparametric estimate of the HRF across all voxels
pertaining to a specific cognitive process. This paper makes several major
contributions with each stated below:
\begin{itemize}
\item MASM constructs a weighted likelihood function by utilizing both
the spatial and frequency information
of fMRI data.
\item The proposed method carries out
a locally adaptive bandwidth selection across different frequencies and
a sequence of
nested spheres with increasing
radii at each voxel to adaptively and spatially estimate HRFs.
\item The estimation procedure uses a back-fitting method to adaptively
estimate HRFs for
multiple stimulus sequences and across all voxels.
\end{itemize}

The rest of the paper is organized as follows. Section \ref{sec2}
presents the
key steps of MASM. Section \ref{sec3}
reports simulation studies to examine the finite sample performance
of MASM. Section \ref{sec4} illustrates an application of MASM in a
real fMRI
data set. Section \ref{sec5} concludes with some discussions.

\section{Model formulation}\label{sec2}

\subsection{Multiscale adaptive smoothing model}\label{sec2.1}

Here we introduce a multiscale adaptive smoothing model for a single
stimulus function. Suppose that we acquire a fMRI data set in a
3D volume, denoted by ${\mathcal D}\subset R^3$, from a
single subject. In the time domain, LTI assumes that for ${\mathbf
d}\in{\mathcal D}$
%
%
\begin{equation}\label{MASMeq1}\qquad
Y(t, {\mathbf d})={H(\cdot, {\mathbf d})}\otimes{X}(t)+\varepsilon(t,
{\mathbf d})=\int{H}(t-u,
{\mathbf d})\cdot{X}(u)\,du + \varepsilon(t, {\mathbf d}),
\end{equation}
where $\otimes$ denotes the circular convolution between two aperiodic
functions and
$\varepsilon(t, {\mathbf d})$ is a measurement error. We observe
$Y(t, {\mathbf d})$ at $T$ acquisition times $t_0,\ldots, t_{T-1}$,
where $t_k=kt_{\mathrm{TR}}$ and $t_{\mathrm{TR}}$ denotes the repetition time, which is
the time between two consecutive scans. Moreover, $T_0=Tt_{\mathrm{TR}}$,
${\mathcal X}=\{X(t)\dvtx t\in[0, T_0]\}$ and ${\mathcal
E}=\{\varepsilon(t, {\mathbf d})\dvtx t\in[0, T_0], {\mathbf d}\in
{\mathcal D}\}$
are assumed to be independent. The error process ${\mathcal E}$ is
assumed to be a stochastic process indexed by $t\in[0, T_0]$ and $
{\mathbf d}\in{\mathcal D}$ with $\mu_\varepsilon(t, {\mathbf
d})=E[\varepsilon(t, {\mathbf d})]=0$ and $\operatorname{Cov}(\varepsilon(t,
{\mathbf
d}), \varepsilon(t', {\mathbf d}'))=\Sigma_T(t, t'; {\mathbf d},
{\mathbf d}')$ for
all $t, t'\in[0, T_0]$ and ${\mathbf d}, {\mathbf d}'\in{\mathcal D}$.
Therefore, the mean function and the covariance function of $Y(t,
{\mathbf d})$ are, respectively, given by
%
%
\begin{eqnarray}
\label{MASMeq21}
E\bigl[Y(t, {\mathbf d})|\mathcal X\bigr] &=&\int{H}(t-u, {\mathbf d})
\cdot{X}(u)\,du,
\\
\label{MASMeq2}
\operatorname{Cov}\bigl(Y(t, {\mathbf d}), Y\bigl(t', {\mathbf d}'
\bigr)\bigr)&=&\Sigma_T\bigl(t, t'; {\mathbf d}, {\mathbf
d}'\bigr).
\end{eqnarray}
In (\ref{MASMeq2}), $\Sigma_T(t, t'; {\mathbf d}, {\mathbf d}')$ can
characterize both temporal and spatial dependence structures in the
fMRI data.




The equivalent model with respect to (\ref{MASMeq1}) in the
frequency domain is obtained by using the Fourier transformation
[\citet{Brillinger1981}, \citet{BrockwellAndDavis1991}, \citet
{FanAndYao2003}]. Let
${\mathcal F}_Y(f, {\mathbf d})$ be the Fourier transformation of $Y(t,
{\mathbf d})$ defined by
%
%
\begin{equation}
\label{MASMeq3} {\mathcal F}_Y (f, {\mathbf d}) =\int
_0^{T_0} Y(t, {\mathbf d}) e^{-2\pi i
ft/T_0} \,dt
\qquad\mbox{for } f\in\Re.
\end{equation}
Similarly, let ${\mathcal F}_H(f, {\mathbf d})$, ${\mathcal F}_X (f)$
and ${\mathcal F}_\varepsilon(f, {\mathbf d})$ be the Fourier
transformations of ${H}(t, {\mathbf d})$, ${X}(t)$ and $\varepsilon(t,
{\mathbf d})$, respectively. In the frequency domain, model
(\ref{MASMeq1}) can be rewritten as
%
%
\begin{equation}
\label{MASMeq4} {\mathcal F}_Y (f, {\mathbf d}) ={\mathcal
F}_H(f, {\mathbf d}){\mathcal F}_X (f)+{\mathcal
F}_\varepsilon(f, {\mathbf d})\qquad\mbox{for } f\in\Re.
\end{equation}
Furthermore, we consider a discrete version of (\ref{MASMeq4}) and
define the discrete Fourier coefficients of $Y(t, {\mathbf d})$, ${H}(t,
{\mathbf d})$, ${X}(t)$ and $\varepsilon(t, {\mathbf d})$ to be, respectively,
$\phi_Y(f_k, {\mathbf d})$, $\phi_H(f_k, {\mathbf d})$, $\phi_X (f_k)$ and
$\phi_\varepsilon(f_k, {\mathbf d})$ at the fundamental frequencies
$f_k=k/T$ for $k=0,\ldots, T-1$. For instance, at $f_k=k/T$, let
$\phi_Y(f_k, {\mathbf d})=\sum_{t=0}^{T-1}\exp{(-2\pi
if_kt)}Y(t,{\mathbf
d})$. Thus, the discrete version of (\ref{MASMeq4}) is given by
%
%
\begin{equation}
\label{MASMeq5} \phi_Y(f_k, {\mathbf d})=
\phi_H(f_k, {\mathbf d}) \phi_{X}(f_k)+
\phi_{\varepsilon}(f_k, {\mathbf d})
\end{equation}
for $k=0, 1,\ldots, T-1$ and all $ {\mathbf d}\in{\mathcal D}$.
Equation (\ref{MASMeq5}) is also a discrete circular convolution.

One advantage of model (\ref{MASMeq5}) in the frequency domain is
that the temporal correlation structure can be substantially
simplified and, thus, the computation burden will be reduced. First,
under some regularity conditions [\citet{ShumwayAndStoffer2006}], the
real and imaginary parts of $\phi_Y(f_k, {\mathbf d})$ are approximately
uncorrelated. Second, if $\varepsilon(t, {\mathbf d})$ is a stationary
error process for each $ {\mathbf d}\in{\mathcal D}$, the Fourier
coefficients are approximately uncorrelated across a pre-specified
set of Fourier frequencies under some regularity conditions
[\citet{BrockwellAndDavis1991}]. Hence, it may be reasonable to assume
ideally that $\phi_{\varepsilon}(f, {\mathbf d})$ is a complex process with
the zero mean function and $\phi_{\varepsilon}(f, {\mathbf d})$ and
$\phi_{\varepsilon}(f', {\mathbf d}')$ are uncorrelated for $f\not=f'$ in
the same voxel ${\mathbf d}={\mathbf d}'$.

Besides the assumptions in (\ref{MASMeq5}),
MASM also assumes two smoothness conditions. The first one is a
frequency smoothness condition. That is, for each $(f, {\mathbf d})\in
[0, 1]\times{\mathcal D}$, there is an open neighborhood of $f$
given the voxel ${\mathbf d}$, denoted by $N_C(f, {\mathbf d})$, such that
$\phi_H(f, {\mathbf d})$ is a continuous function of $f$. 
The first condition allows us to consistently estimate
$\phi_H(f, {\mathbf d})$ by solely using the data in voxel~$\mathbf{d}$.
The second one is a joint frequency and spatial smoothness
condition. Specifically, there is a frequency-spatial neighborhood
of $(f, {\mathbf d})$, denoted by $N_J(f, {\mathbf d})$, such that there
exists at least a sequence $\{(f_n, {\mathbf d}_n)\}$ in $N_J(f,
{\mathbf
d})$ which satisfies
%
%
\begin{equation}
\label{MASMeq6} \lim_{n\rightarrow\infty}(f_n, {\mathbf
d}_n)= ( f, {\mathbf d}) \quad\mbox{and}\quad \lim_{n\rightarrow\infty}
\phi_H(f_n, {\mathbf d}_n)=\phi_H(
f, {\mathbf d}).
\end{equation}
The set $N_J(f, {\mathbf d})$ is always nonempty, since it at least
contains $N_C(f, {\mathbf d})$ given that $(f,{\mathbf d})\in
N_C(f,{\mathbf
d})$. The second condition allows us to incorporate fMRI data in a
frequency-spatial neighborhood of $(f, {\mathbf d})$ to estimate
$\phi_H(f, {\mathbf d})$. Assumption (\ref{MASMeq6}) may be reasonable
for the
real fMRI data since the fMRI data often contain spatially
contiguous homogenous regions with rather sharp edges. When ${\mathbf
d}$ varies in ${\mathcal D}$, assumption (\ref{MASMeq6}) allows for
neighborhoods with varying
shapes across the 3D volume, and thus it can characterize varying
degrees of spatial
smoothness. Moreover, under (\ref{MASMeq6}), MASM essentially treats
$\{\phi_Y(f, {\mathbf d})\}$ as a stochastic process indexed
by both frequency $f$ and voxel $ {\mathbf d}$.

\subsection{Weighted least square estimate}\label{sec2.2}

Our goal is to estimate the unknown function $\{\phi_H(f, {\mathbf
d})\dvtx
f\in[0, 1], {\mathbf d}\in{\mathcal D}\}$ in MASM defined in
(\ref{MASMeq5}) and (\ref{MASMeq6}) based on the Fourier transformed
fMRI data $ {\mathcal F}({\mathbf Y})=\{\phi_Y(f_k, {\mathbf
d})\dvtx k=0,\ldots, T-1, {\mathbf d}\in{\mathcal D}\}$. To estimate
$\phi_H(f, {\mathbf d})$, we combine the data at $(f_k=k/T, {\mathbf d}')$
near $(f, {\mathbf d})$ to set up an approximation equation as follows:
%
%
\begin{eqnarray}\label{MASMeq7}
\phi_Y\bigl(f_k, {\mathbf d}'\bigr)&=&
\phi_H\bigl(f_k, {\mathbf d}'\bigr)
\phi_{X}(f_k)+\phi_{\varepsilon}\bigl(f_k, {
\mathbf d}^{\prime}\bigr)
\nonumber\\[-8pt]\\[-8pt]
&\approx& \phi_H(f, {\mathbf d}) \phi_{X}(f_k)+
\phi_{\varepsilon}\bigl(f_k, {\mathbf d}^{\prime}\bigr).
\nonumber
\end{eqnarray}

Based on model (\ref{MASMeq7}), we can construct a weighted function
at $(f, {\mathbf d})$. For simplicity, we consider all
$f_k\in I(f,r)=(f-r, f+r)$ and all $ {\mathbf d}^{\prime}\in B({\mathbf d},
h)$, where $r>0$ and $B({\mathbf d},h)$ is a spherical neighborhood of
voxel ${\mathbf d}$ with radius $h\ge0$. Then, to estimate $\phi_H(f,
{\mathbf d})$, we construct a locally weighted function, denoted as
$L(\phi_H(f, {\mathbf d}); r, h)$, which is given by
%
%
\begin{equation}
\label{MASMeq8}\quad \sum_{f_k\in I(f,r)}\sum
_{{\mathbf d}^{\prime}\in B({\mathbf d}, h)}\bigl|\phi_{Y
}\bigl(f_k, {\mathbf
d}^\prime\bigr)-\phi_H(f, {\mathbf d})\phi_{
X}(f_k)\bigr|^2{
\tilde{\omega}\bigl({\mathbf d}, {\mathbf d}^{\prime},f, f_k; r, h
\bigr)},
\end{equation}
where $|\cdot|$ denotes the norm of a complex number. Moreover,
${\tilde{\omega}({\mathbf d}, {\mathbf d}^{\prime},f, f_k; r, h)}$ is a
nonnegative weight function such that
%
%
\begin{equation}
\sum_{f_k\in
I(f,r)}\sum_{{\mathbf d}^{\prime}\in B({\mathbf d},h)}
\tilde{\omega}\bigl({\mathbf d},{\mathbf d^{\prime}},f,f_k;r,h\bigr)=1
\end{equation}
for all ${\mathbf d}\in
{\mathcal D}$ and $f\in[0, 1]$. The right choice of
${\tilde{\omega}({\mathbf d}, {\mathbf d}^{\prime},f, f_k; r, h)}$ in
(\ref{MASMeq8}) is the key to the success of MASM.
In Section \ref{sec2.3} we explicitly define all weights $\tilde
{\ve{\omega}}(r, h)=\{{\tilde{\omega}({\mathbf d}, {\mathbf d}^{\prime},f,
f_k; r, h)}\dvtx {\mathbf d}, {\mathbf d}'\in{\mathcal D}, f, f_k\in
[0, 1]\}$
for the fixed $r$ and $h$.


Given ${\tilde{\ve{\omega}}(r, h)}$, by differentiating $L(\phi_H(f,
{\mathbf d}); r, h)$ with respect to $\phi_{H}(f, {\mathbf d})$, we have
%
%
\begin{equation}
\label{MASMeq9} \hat\phi_{H}(f, {\mathbf d})={\sum_{f_k\in I(f,r)}\sum
_{{\mathbf
d}^{\prime}\in B({\mathbf d},h)}{\tilde{\omega}({\mathbf d}, {\mathbf
d}^{\prime},f, f_k; r, h)}\overline{\phi_{ X}(f_k)}\phi_Y(f_k, {\mathbf
d}^{\prime})\over\sum_{f_k\in I(f,r)}\sum_{{\mathbf d}^{\prime}\in
B({\mathbf d}, h)}{\tilde{\omega}({\mathbf d}, {\mathbf d}^{\prime},f,
f_k; r,
h)}\phi_{ X}(f_k)\overline{\phi_{X}(f_k)}},\hspace*{-34pt}
\end{equation}
where $\overline{\phi_{X}(f_k)}$ is the conjugate of $ {\phi_{
X}(f_k)}$. The variance of $\hat\phi_{H}(f, {\mathbf d})$ is
approximated by
%
%
\begin{eqnarray}\label{MASMeq10}
&&
\operatorname{Var}\bigl(\hat\phi_{H}(f, {\mathbf d})\bigr)\nonumber\\
&&\qquad\approx E
\bigl[\bigl\{\hat{\phi}_H(f,{\mathbf d})-\phi_H(f,{\mathbf d})
\bigr\}\overline{\bigl\{\hat{\phi}_H(f,{\mathbf d})-
\phi_H(f,{\mathbf d})\bigr\}}\bigr]
\\
&&\qquad\approx{\sum_{f_k\in I(f,r)}|{\sum_{{\mathbf d}^{\prime}\in B({\mathbf d},
h)}}\tilde{\omega}({\mathbf d}, {\mathbf
d}^{\prime},f,f_k;r,h)\overline{\phi_X(f_k)}\hat{\phi}_{\varepsilon}(f_k,
{\mathbf d}^{\prime})|^2 \over\{\sum_{f_k\in I(f,r)}\sum_{{\mathbf
d}^{\prime}\in B({\mathbf d}, h)}{\tilde{\omega}({\mathbf d}, {\mathbf
d}^{\prime},f, f_k; r,
h)}\phi_{X}(f_k)\overline{\phi_{X}(f_k)}\}^2},
\nonumber
\end{eqnarray}
where $\hat\phi_\varepsilon(f_k, {\mathbf d}^{\prime})=\phi_Y(f_k,
{\mathbf
d}')-\hat\phi_H(f_k, {\mathbf d}') \phi_{X}(f_k)$ and
the last approximation is based on the de-correlation between two
different Fourier frequencies.

By taking the inverse Fourier transformation of $\hat\phi_{H}(f,
{\mathbf d})$, we can derive
%
%
\begin{equation}
\label{MASMeq11} \tilde{H}(t, {\mathbf d})={1\over
T}\sum
_{k=0}^{T-1}\hat\phi_{H}(f_k, {
\mathbf d})\exp{(i2\pi tf_k)}
\end{equation}
for any $ {\mathbf d}\in{\cal D}$ and $t$. As discussed in
\citet{Brillinger1974} and \citet{Bohman1960}, since the whole real
domain in the
Fourier transformation is truncated to the domain $[0,T]$,
the estimator of $H(t, {\mathbf d})$ by directly using the inverse Fourier
transformation can be biased. Therefore, we use a tapered estimator of
$H(t, {\mathbf d})$ as follows:
%
%
\begin{equation}
\label{MASMeq12}\qquad \hat{H}(t, {\mathbf d})=\sum_{k=0}^{T-1}
\hat\phi_{H}(f_k, {\mathbf d})\exp{(i{2\pi
tf_k})}\biggl[{1-\cos\biggl({2\pi\over T}t\biggr)\biggr]\Big/
\biggl[\pi{2\pi\over T}t^2}\biggr].
\end{equation}

\subsection{Multiscale adaptive estimation procedure}\label{sec2.3}

We use a multiscale adaptive estimation (MAE) procedure to determine
all weights $\tilde{\ve{\omega}}(r, h)$ and then estimate
$\{\phi_H(f, {\mathbf d})\dvtx {\mathbf d}\in{\mathcal D}, f\in[0, 1]\}
$. MAE
extends the multiscale adaptive strategy from the
propagation--separation (PS) approach
[Polzehl and Spokoiny (\citeyear{PolzehlAndSpokoiny2000}, \citeyear{PolzehlAndSpokoiny2006})]. MAE starts
with building two sequences at each $(f, {\mathbf d})\in[0, 1]\times
{\mathcal D}$. The first is a sequence of nested spheres denoted by
%
%
\begin{equation}\hspace*{28pt}
B({\mathbf d}, h_0)\subset\cdots\subset B({\mathbf d}, h_S)
\qquad\mbox{for increasing } h_0=0<h_1<\cdots<h_S.
\end{equation}
Increasing the spatial radius $h$, from the smallest scale $h_0=0$
to the largest scale $h_S$ at each $d\in{\mathcal D}$, allows us to
control the degree of smoothness in the spatial domain. The second
one is a sequence of nested intervals given by
%
%
\begin{equation}\hspace*{28pt}
I(f, r_0)\subset\cdots\subset I(f, r_S)\qquad\mbox{for
increasing } 0<r_0<r_1<\cdots<r_S.
\end{equation}
Increasing the frequency radius $r$ from some smallest scale $r_0>0$
to the largest scale $r_S$ at each $f\in[0, 1]$ allows us to
control the degree of smoothness in the frequency domain. After
calculating $\tilde{\ve{\omega}}(r_0, h_0)$, we can estimate
$\phi_H(f, {\mathbf d})$ at the smallest scale $(r_0, h_0)$, denoted by
$\hat\phi_H^{(0)}(f, {\mathbf d})$. Then, based on the information
contained in $\hat\phi_H^{(0)}=\{\hat\phi_H^{(0)}(f, {\mathbf d})\dvtx
{\mathbf d}\in{\mathcal D}, f\in[0, 1]\}$, we use the methods
described below to calculate a set of weights $\tilde
{\ve{\omega}}(r_l, h_l)$ at radii $h_l$ and $r_l$ for all $(f, d)\in
[0,1]\times{\mathcal D}$. 
Sequentially, 
we update the estimates $\hat\phi_H^{(l)}=\{\hat\phi_H^{(l)}(f,
{\mathbf d})\dvtx {\mathbf d}\in{\mathcal D}, f\in[0, 1]\}$ according to
(\ref{MASMeq9}). At each iteration, we also calculate a stopping
test statistic at each ${\mathbf d}\in{\cal D}$, denoted as $W({\mathbf d};
h_l, r_l)$, to prevent over-smoothing $\{\phi_H(f,{\mathbf d})\dvtx
f\in[0,1]\}$.

The MAE procedure consists of four key steps: (i) initialization,
(ii) weights adaptation, (iii) estimation, and (iv) stop checking.
These steps are presented as follows:
\begin{itemize}
\item\textit{Initialization}. In this step we set $h_0=0$, $r_0>0$, say,
$r_0=5/T$, and the weighting scheme
$\tilde{\omega}({\mathbf d}, {\mathbf d}, f, f_k; r_0,
h_0)=K_{\mathrm{loc}}(|f-f_k|/r_0)$, where $K_{\mathrm{loc}}(x)$ is a kernel function
with compact support. Then we\vspace*{1pt} substitute $\tilde{\omega}({\mathbf d},
{\mathbf d}, f,
f_k;\break r_0, u_0)$ into (\ref{MASMeq9}) and (\ref{MASMeq10}) to
calculate $\hat\phi_{H}^{(0)}(f, {\mathbf d})$ and estimate\break
$\operatorname{Var}(\hat\phi_{H}^{(0)}(f, {\mathbf d}))$. We also set
up a
geometric series $\{h_l=c_h^l\dvtx l=1,\ldots, S\}$ for the spatial
radii, where $c_h\in(1,2)$, say, $c_h=1.125$, and then we set up the
second series $\{r_l=r_{l-1}+b_r\dvtx l=1,\ldots, S\}$ for the
frequency radii, where $b_r$ is a constant value, say, $b_r=1/T$.
We set $l=1$ and $h_1=c_h$.

\item\textit{Weight adaptation}. In this step we compute the
adaptive weight $\tilde{\omega}({\mathbf d}, {\mathbf d}^{\prime},\allowbreak f,
f_k;
r_l, h_l)$, which is given by
%
%
\begin{eqnarray}
\label{MASMeq14}
&&
K_{\mathrm{loc}}\bigl(\bigl\|{\mathbf d}-{\mathbf d}^{\prime}\bigr\|_2/h_l
\bigr) K_{\mathrm{loc}}\bigl(|f-f_k|/r_l\bigr)\nonumber\\[-8pt]\\[-8pt]
&&\qquad{}\times K_{st}
\biggl({|\hat\phi_{H}^{(l-1)}(f, {\mathbf
d})-\hat\phi_{H}^{(l-1)}(f_k, {\mathbf
d}^{\prime})|\over{\sqrt{\operatorname{Var}(\hat\phi_{H}^{(l-1)}(f,
{\mathbf
d}))}}} \biggr),\nonumber
\end{eqnarray}
where \mbox{$\|\cdot\|_2$} is the Euclidean norm. 
The functions $K_{\mathrm{loc}}(x)$ and $K_{st}(x)$ are two kernel functions
within compact support. The
$\operatorname{Var}(\hat\phi_{H}^{(l-1)}(f, {\mathbf d}))$ is the estimated
variance of $\hat\phi_{H}^{(l)}(f, {\mathbf d})$ at the $(l-1)$th step.
See (\ref{MASMeq10}) for details. The $K_{st}(x)$
downweights the information at $( f_k, {\mathbf d}^{\prime})$ for large
$\|\hat\phi_{H}^{(l-1)}(f, {\mathbf d})-\hat\phi_{H}^{(l-1)}(f_k,
{\mathbf
d}^{\prime})\|$. The first two kernel functions give less weight to
$(f_k, {\mathbf d}^{\prime})$, which is far from $(f, {\mathbf d})$.

\item\textit{Estimation}. In this step we substitute the weight $\tilde
{\omega}({\mathbf d}, {\mathbf d}^{\prime}, f, f_k; r_l, h_l)$ into
(\ref{MASMeq9}) and (\ref{MASMeq10}) in order to calculate
$\hat\phi_{H}^{(l)}(f_k, {\mathbf d})$ and
$\operatorname{Var}(\hat\phi_{H}^{(l)}(f_k, {\mathbf d}))$ at the $k$th
fundamental frequency $f_k$ and voxel ${\mathbf d}$.

\item\textit{Stop checking}. In this step, after the $S_0$th iteration
for some $S_0>0$ and $S_0<S$, we calculate a stopping criterion based
on a normalized distance between
$\hat\phi_{H}^{(l)}({\mathbf d})=\{\hat\phi_{H}^{(l)}(f_0, {\mathbf
d}),\ldots, \hat\phi_{H}^{(l)}(f_{T-1}, {\mathbf d})\}$ and\vspace*{1pt} $\hat\phi
_{H}^{(l-1)}({\mathbf d})=\{\hat\phi_{H}^{(l-1)}(f_0, {\mathbf
d}),\ldots,\allowbreak \hat\phi_{H}^{(l-1)}(f_{T-1}, {\mathbf d})\}$.
Specifically, we calculate a test statistic $W^{(l)}({\mathbf d};
h_l,r_l)$ to test the following hypotheses:
\[
H_{N}\dvtx \hat\phi_{H}^{(l)}({\mathbf d})-\hat
\phi_{H}^{(l-1)}({\mathbf d})={\mathbf0} \quad\mbox{versus}\quad
H_{A}\dvtx \hat\phi_{H}^{(l)}({\mathbf d})-\hat
\phi_{H}^{(l-1)}({\mathbf d})\not={\mathbf0}.
\]
The $W^{(l)}({\mathbf d};
h_l, r_l)$ is an adaptive Neyman test statistic for testing the
potential difference between the two frequency series
[\citet{FanAndHuang2001}]. See the explicit form of
$W^{(l)}({\mathbf d};
h_l, r_l)$ in Part A of the supplementary material [\citet
{Jiaetal}]. If $W^{(l)}({\mathbf d}; h_l, r_l)$
is significant at a given significance level $\alpha$, say, 0.05,
then we set $\hat\phi^{(S)}_H(f, {\mathbf d})=\hat\phi_H^{(l-1)}(f,
{\mathbf d})$ and $l=S$ at voxel ${\mathbf d}$. If $l=S$ for all
${\mathbf d}\in
{\mathcal D}$, we stop the MAE procedure. If $l\le S$ or
$W^{(l)}({\mathbf d}; h_l,r_l)$ is not significant, then we set $
h_{l+1}=c_h h_{l}$ and $r_{l+1}=r_l+b_r$, increase $l$ by 1, and
continue with the weight adaptation step (ii).
\end{itemize}

Finally, we report the final $\hat\phi^{(S)}_H(f, {\mathbf d})$ at
all fundamental frequencies and substitute them into
(\ref{MASMeq12}) to calculate $\hat{H}(t, {\mathbf d})$ for all voxels
${\mathbf d}\in{\mathcal D}$.


\begin{Remarks*} The finite-sample performance of the MAE procedure
depends on the specification of some key parameters, including $S$,
$c_h$, $S_0$, $b_0$, $b_r$ and the kernel functions $K_{\mathrm{loc}}(\cdot)$
and $K_{st}(\cdot)$. We have tested different combinations of key
parameters in both simulated and real data. The performance of MAE is
quite robust to moderate changes in $c_h$, $S_0$, $b_0$, $b_r$ and $S$.
See Part C of the supplementary material [\citet{Jiaetal}] for
additional
simulations.

For the kernel functions, we set
\begin{eqnarray*}
K_{\mathrm{loc}}(x)&=&\bigl(1-x^2\bigr){\mathbf1}(x\le1),
\\
K_{st}(x)&=& 1-6x^2+6x^3{\mathbf1}\bigl(x\in[0,
0.5]\bigr)+2(1-x)^3{\mathbf1}(x\in(0.5,1]).
\end{eqnarray*}
The latter one is the Parzen
window [\citet{FanAndYao2003}]. Other
choices of the kernel functions include the kernel functions in the
original PS approach [Polzehl and Spokoiny (\citeyear{PolzehlAndSpokoiny2000}, \citeyear{PolzehlAndSpokoiny2006}), Tabelow
et al. (\citeyear{TabelowEtal2006}, \citeyear{TabelowEtal2008})]
or the Gaussian kernels. Since the initial estimators of $\phi_H(f,
{\mathbf d})$ are solely calculated in the frequency domain, they are
pretty robust to the choice of kernel function but
sensitive to the bandwidth selection. So we select a small
bandwidth, say, $5/T$, as the initial value, and then we use
the adaptive procedure to determine a better
estimation by slowly increasing the bandwidths of $(f, {\mathbf
d})$.\eject

The parameters $h_l$ and $r_l$ play the same role as the bandwidth
of local kernel methods. The small values of $h_l$ and $r_l$ only
incorporate the closest neighboring voxels and the closest
frequencies of $(f,{\mathbf d})$. Thus, small
values of $c_h$ and $b_r$ can prevent over-smoothing
$\hat\phi_{H}(f, {\mathbf d})$ at the beginning of MAE and improve
the robustness of MAE, whereas small
values of $c_h$ and $b_r$ lead to increased computational effort.
We have found that $c_h=1.125$ and $b_r=1/T$ perform well in numerous
simulations.

We suggest to set $S_0$ as a small integer, say, 2 or 3.
Large values of $S_0$ lead to both heavy computation and over-smoothing
when a voxel ${\mathbf d}$ is either on the boundary of significant regions
or in some regions in which the HRFs change
slowly with voxel location. After the $S_0$th iteration, the stop
checking step starts to compute the stopping criterion and check
whether further iteration is needed in this voxel. Moreover, the stop
checking step is essentially a bandwidth selection procedure. In the
original PS approach [Polzehl and Spokoiny (\citeyear{PolzehlAndSpokoiny2000}, \citeyear{PolzehlAndSpokoiny2006}), Tabelow
et al. (\citeyear{TabelowEtal2006}, \citeyear{TabelowEtal2008})],
a Wald-type statistic was used to compare consecutive
estimates in order to prevent over-smoothing the parameters in the
estimated images. Since HRF is an infinite-dimensional function, we
employ the adaptive Neyman test statistic
[\citet{FanAndHuang2001}]. Actually, our stop checking step is to set
some tolerance (e.g., significance level) and iterate until the
difference is less than that threshold and, thus, it is the same as
that used in original PS approach [Polzehl and Spokoiny (\citeyear{PolzehlAndSpokoiny2000}, \citeyear{PolzehlAndSpokoiny2006}),
Tabelow et al. (\citeyear{TabelowEtal2006}, \citeyear{TabelowEtal2008})].

As the maximal iteration $S$ increases, the number of neighboring
voxels in $B({\mathbf d}, c_h^S)$ increases exponentially and the
number of time points in $I(f, r_l)$ increases linearly. Moreover, a
large $S$ also increases the probability of over-smoothing
$\phi_H({\mathbf d})$ when the current voxel ${\mathbf d}$ is near the edge
of distinct regions and the HRFs change slowly with other locations.
In practice, we suggest the maximal step $S$ up to be 15 but larger
than 10.

Although spherical neighborhoods lead to a computationally simple
procedure, the use of these spherical neighborhoods is not an
optimal way of incorporating imaging data in ``good'' voxels, which
are close to the imaging data in the current voxel. Particularly,
for those voxels near the boundary of activated regions, spherical
neighborhoods may cover many ``bad'' voxels. It is interesting
to determine multiscale neighborhoods adaptive to the
pattern of imaging data at each voxel, which is our ongoing research.
\end{Remarks*}

\subsection{Multiple stimuli}\label{sec2.4}

\begin{figure}

\includegraphics{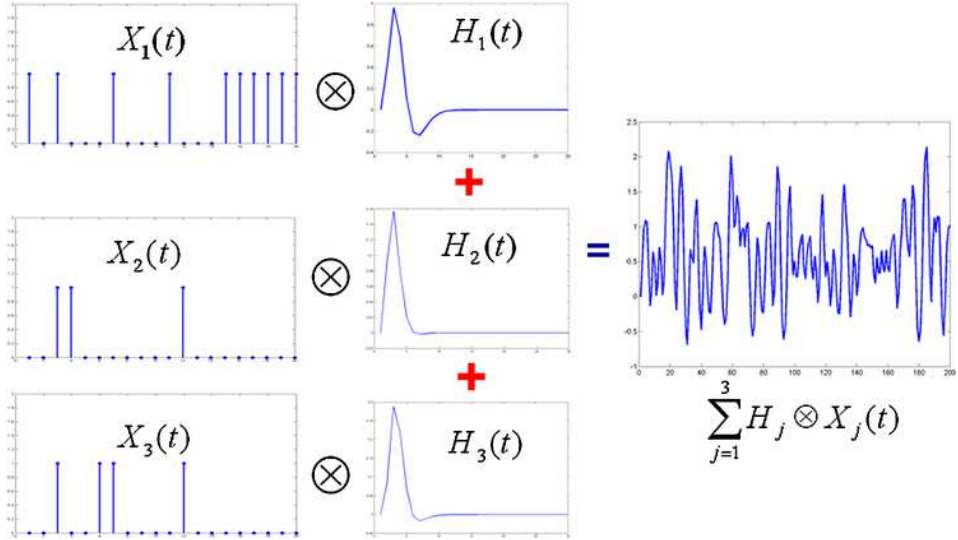}

\caption{A diagram of the case with multiple stimuli. The fMRI signals
are the sum of three HRFs convoluted with the corresponding
sequences of stimulus events. The $X_j(t)$ is the stimulus sequence and
$H_j(t)$ is the HRF for $j=1,2$ and $3$. We ignore the voxel ${\mathbf
d}$ in
$X_j(t)$ and $H_j(t)$ for notational simplicity.} \label{figure3}
\end{figure}

In the real fMRI studies, it is common that multiple stimuli are
present. In this section we generalize MASM to the case of multiple
stimuli. Under the assumption of the LTI system, the BOLD signal is the
sum of the individual responses to all stimuli convoluted with their
associated HRFs. See a sample path diagram in Figure \ref{figure3}. Let
%
%
${\mathbf X}(t)=(X_1(t),\ldots, X_m(t))^{\mathrm{T}}$ be the sequence
vector of $m$ different stimuli and its associated HRF vector ${\mathbf
H}(t, {\mathbf d})=(H_1(t, {\mathbf d}),\ldots, H_m(t, {\mathbf
d}))^{\mathrm{T}}$. Specifically, in the time domain, our MASM under
the presence of $m$ different stimuli is given by
%
%
\begin{equation}
\label{MASMeq15} Y(t, {\mathbf d})=\int\bigl\langle{\mathbf H}({\mathbf d}, t-u),
{\mathbf X}(u)\bigr\rangle\,du +
\varepsilon({\mathbf d}, t),
\end{equation}
%
where $\langle\cdot,\cdot\rangle$ is an inner product defined as
$\langle A,B\rangle=A^{\mathrm{T}}B$ for two vectors $A$ and $B$. Subsequently, in the
frequency domain, the discrete version of MASM for multiple stimuli
is given by
%
%
\begin{equation}
\label{MASMeq16} \phi_{Y}(f, {\mathbf d})=\bigl\langle\phi_{\mathbf H}(f, {\mathbf
d}), \phi_{\mathbf
X}(f)\bigr\rangle+\phi_{\varepsilon}(f, {\mathbf d}),
\end{equation}
where
%
%
\begin{eqnarray}
\phi_{\mathbf H}(f, {\mathbf d})&=& \bigl(\phi_{ H_1}(f, {\mathbf
d}),\ldots,
\phi_{ H_m}(f, {\mathbf d})\bigr)^{\mathrm{T}},
\\
%
\phi_{\mathbf X}(f)&=& \bigl(\phi_{X_1}(f),\ldots,
\phi_{X_m}(f)\bigr)^{\mathrm{T}}.
\end{eqnarray}

Furthermore, the MASM for multiple stimuli assumes that for each $j$,
$\phi_{H_j}(f,{\mathbf d})$ satisfies both the
frequency smoothness condition in an open neighborhood of $f$,
denoted as $N_{C_j}(f,{\mathbf d})$, and the joint frequency and spatial
smoothness condition in $N_{J_j}(f,{\mathbf d})$, a neighborhood of
$(f,{\mathbf d})$. Note that $N_{C_j}(f,{\mathbf d})$ and
$N_{J_j}(f,{\mathbf d})$ may vary across different $j$,
since HRFs vary across
different $j$ and cognitive processes. In this case, one
cannot use the same weights $\tilde{\omega}({\mathbf d}, {\mathbf
d}^{\prime}, f, f_k;r, h )$ for all stimuli since different stimuli
may have different degrees of smoothness near each $({\mathbf d}, f)$.
We present an alternative approach below.

We construct $m$ locally weighted functions, denoted as
$L_{j}(\phi_{H_j}(f, {\mathbf d}); r, h)$, given by
%
%
\begin{eqnarray}
\label{MASMeq18}
&&\sum_{f_k\in I(f,r),
{\mathbf d}^{\prime}\in B({\mathbf d},
h)}\bigl|\phi_{Y[-j]}
\bigl(f_k, {\mathbf d}^\prime\bigr)-\phi_{H_j}(f, {\mathbf d})\phi_{
X_j}(f_k)\bigr|^2\nonumber\\[-8pt]\\[-8pt]
&&\hspace*{74pt}{}\times{\tilde{
\omega}_j\bigl({\mathbf d}, {\mathbf d}^{\prime},f, f_k;
r, h\bigr)}\nonumber
\end{eqnarray}
for $j=1,\ldots, m$, where $\phi_{Y[-j]}(f_k, {\mathbf
d}^\prime)=\phi_{Y}(f_k, {\mathbf d}^\prime)-\sum_{l\not=j}\phi
_{H_{l}}(f_k, {\mathbf d}')\phi_{ X_j}(f_k)$. Moreover,
${\tilde{\omega}_j({\mathbf d}, {\mathbf d}^{\prime},f, f_k; r, h)}$
characterizes the physical distance between $(f, {\mathbf d})$ and
$(f_k, {\mathbf d}^{\prime})$ and the similarity between $\phi_{H_j}(f,
{\mathbf d})$ and $\phi_{H_j}(f_k, {\mathbf d}^{\prime})$. Similar to
(\ref{MASMeq9}) and (\ref{MASMeq10}), we can derive recursive formula
to update $\phi_{H_j}(f, {\mathbf d})$ and
$\operatorname{Var}(\hat\phi_{H_j}(f, {\mathbf d}))$ for $j=1,\ldots,
m$ based on any fixed weights $\{{\tilde{\omega}_j({\mathbf d},
{\mathbf d}^{\prime},f, f_k; r, h)}\dvtx\allowbreak {\mathbf
d}^{\prime}\in B({\mathbf d}, h), f_k\in I(f,r)\}$. By differentiating
$L_j(\phi_{H_j}(f, {\mathbf d});\break r, h)$ with respect to $\phi_{H_j}(f,
{\mathbf d})$, we can get
%
%
\begin{eqnarray}
\label{MASMeq19}
&&
\hat\phi_{H_j}(f, {\mathbf d})\nonumber\\[-8pt]\\[-8pt]
&&\qquad={\sum_{f_k\in
I(f,r)}\sum_{{\mathbf d}^{\prime}\in B({\mathbf d},h)}{\tilde{\omega
}_j({\mathbf d}, {\mathbf d}^{\prime},f, f_k; r, h)}\overline{\phi_{
X_j}(f_k)}\phi_{Y[-j]}(f_k, {\mathbf d}^{\prime})\over\sum_{f_k\in
I(f,r)}\sum_{{\mathbf d}^{\prime}\in B({\mathbf d},
h)}{\tilde{\omega}_j({\mathbf d}, {\mathbf d}^{\prime},f, f_k; r,
h)}\phi_{
X_j}(f_k)\overline{\phi_{X_j}(f_k)}}.\hspace*{-28pt}\nonumber
\end{eqnarray}
%
Then, we approximate the variance of $\hat\phi_{H_j}(f, {\mathbf d})$,
denoted as $\operatorname{Var}(\hat\phi_{H_j}(f, {\mathbf d}))$, as follows:
%
%
\begin{eqnarray}\label{MASMeq20}
&&{E[|\sum_{f_k\in
I(f,r)}\sum_{{\mathbf d}^{\prime}\in B({\mathbf d}, h)}\tilde{\omega
}_j({\mathbf d}, {\mathbf d}^{\prime},f,f_k;r,s)\overline{\phi
_{X_j}(f_k)}\phi_\varepsilon(f_k,
{\mathbf d}^{\prime})|^2]\over\{\sum_{f_k\in I(f,r)}\sum_{{\mathbf
d}^{\prime}\in B({\mathbf d}, h)}{\tilde{\omega}_j({\mathbf d},
{\mathbf d}^{\prime},f, f_k; r, h)}\phi_{
X_j}(f_k)\overline{\phi_{X_j}(f_k)}\}^2}
\nonumber\\[-8pt]\\[-8pt]
&&\qquad\approx{\sum_{f_k\in I(f,r)}|\sum_{{\mathbf d}^{\prime}\in B({\mathbf d},
h)}\tilde{\omega}_j({\mathbf d}, {\mathbf d}^{\prime
},f,f_k;r,h)\overline{\phi_{X_j}(f_k)}\hat\phi_\varepsilon(f_k,
{\mathbf d}^{\prime})|^2\over\{\sum_{f_k\in I(f,r)}\sum_{{\mathbf
d}^{\prime}\in B({\mathbf d}, h)}{\tilde{\omega}_j({\mathbf d},
{\mathbf d}^{\prime},f, f_k; r, h)}\phi_{
X_j}(f_k)\overline{\phi_{X_j}(f_k)}\}^2},\hspace*{-28pt}
\nonumber
\end{eqnarray}
where $\hat\phi_\varepsilon(f_k, {\mathbf d}^{\prime})
=\phi_{Y}(f_k, {\mathbf d}^\prime)-\sum_{j=1}^m\hat\phi_{H_{j}}(f_k,
{\mathbf d}')\phi_{
X_j}(f_k)$.




Based on the discussions above, we can develop an MAE procedure for
multiple stimuli. The key idea of MAE for multiple stimuli is to
integrate MAE for the single stimulus sequence and the backfitting
method [\citet{BreimanAndFriedman1985}]. Thus, it can sequentially and
recursively compute $\hat\phi_{H_j}(f, {\mathbf d})$ as $j$ increases
from~1 to $m$. For the sake of space, we highlight several key
differences between MAE for a single stimulus and MAE for multiple
stimuli. Generally, MAE consists of four key steps: (i)
initialization, (ii) weight adaption, (iii) recursive estimation, and
(iv)~stopping check.

\begin{itemize}
\item\textit{Initialization}. We use the backfitting method
[\citet{BreimanAndFriedman1985}] to iteratively compute
$\hat\phi_{H_j}^{(0)}(f, {\mathbf d})$ and estimate
$\operatorname{Var}(\hat\phi_{H_j}^{(0)}(f, {\mathbf d}))$ based on
$\phi_{Y[-j]}(f,\allowbreak{\mathbf d})\approx\phi_{H_j}(f,{\mathbf d})\phi
_{X_j}(f)+\phi_{\varepsilon}(f,{\mathbf d})$ for $j=1,\ldots, m$.


\item\textit{Weight adaptation}. We compute $\tilde{\omega
}_j^{(l)}({\mathbf d}, {\mathbf d}^{\prime}, f, f_k;r_l, h_l)$ as follows:
%
%
\begin{eqnarray}
\label{MASMeq21}
&&
K_{\mathrm{loc}}\bigl(\bigl\|{\mathbf d}-{\mathbf d}^{\prime}\bigr\|_2/h_l
\bigr) K_{\mathrm{loc}}\bigl(|f-f_k|/r_l\bigr)\nonumber\\[-8pt]\\[-8pt]
&&\qquad{}\times K_{st}
\biggl({|\hat\phi_{H_j}^{(l-1)}(f, {\mathbf d})-\hat\phi
_{H_j}^{(l-1)}(f_k, {\mathbf d}^{\prime})|\over\sqrt{\operatorname
{Var}(\hat\phi_{H_j}^{(l-1)}(f, {\mathbf d}))}} \biggr).\nonumber
\end{eqnarray}

\item\textit{Recursive estimation}. At the $l${th} iteration, we compute
$\phi_{Y[-j]}^{(l)}(f, {\mathbf d})=\phi_Y(f,\allowbreak {\mathbf d})-\sum_{l\not
=j}\hat\phi_{H_l}^{(l-1)}(f, {\mathbf d}) \phi_{X_l}(f)$.
Then, based on weights $\tilde{\omega}_j^{(l)}({\mathbf d}, {\mathbf
d}^{\prime}, f, f_k;\break r_l, h_l)$, we use the backfitting method
[\citet{BreimanAndFriedman1985}] to iteratively calculate
$\hat\phi_{H_j}^{(l)}(f, {\mathbf d})$ and approximate
$\operatorname{Var}(\hat\phi_{H_j}^{(l)}(f, {\mathbf d}))$ according to
(\ref{MASMeq19}) and (\ref{MASMeq20}).

\item\textit{Stop checking}. After the $S_0$th iteration, we calculate
the adaptive Neyman test statistic, denoted by $W^{(l)}_j({\mathbf d};
h_l,r_l)$, to test difference between
$\hat\phi_{H_j}^{(l)}({\mathbf d})=\{\hat\phi_{H_j}^{(l)}(f_0, {\mathbf
d}),\ldots, \hat\phi_{H_j}^{(l)}(f_{T-1}, {\mathbf d})\}$ and $\hat\phi
_{H_j}^{(l-1)}({\mathbf d})=\{\hat\phi_{H_j}^{(l-1)}(f_0, {\mathbf
d}),\ldots,\break \hat\phi_{H_j}^{(l-1)}(f_{T-1}, {\mathbf d})\}$ for the
$j$th stimulus.
\end{itemize}

Finally, when $l=S$, we report the final $\hat\phi_{H_j}^{(S)}(f,
{\mathbf d})$ at all fundamental frequencies and substitute them into
(\ref{MASMeq12}) to calculate $\hat{H}_j(t, {\mathbf d})$ across voxels
${\mathbf d}\in{\mathcal D}$ for $j=1,\ldots, m$.

After obtaining HRFs for all stimuli, we may calculate their summary statistics,
including $H_a$, $T_p$ and
$W$, and then carry out statistical
inference based on the images of these estimated summary statistics.
For instance, most fMRI studies focus on comparing the $H_a$ images
across diagnostic groups or across stimuli by using voxel-wise
methods [\citet{LindquistAndWager2007}]. Specifically, the voxel-wise
methods involve fitting a statistical model, such as a linear model,
to HRF summary data from all subjects at each voxel and generating a
statistical parametric map of test statistics and $p$-values
[\citet{NicholsAndHolmes2002}, \citet{WorsleyEtal2004},
\citet{ZhangEtal2011}].

\section{Simulation studies}\label{sec3}

We conducted two sets of simulation studies to examine the finite
sample performance of MASM and MAE and compared them with several
state-of-the-art models for estimating HRFs. To present the results
clearly, we also implemented an EM-based algorithm to cluster the
estimated HRFs, which is described in Part B of the supplementary
material [\citet{Jiaetal}], and will present it in a companion paper.

\subsection{Simulation I: One stimulus sequence}\label{sec3.1}

\begin{figure}

\includegraphics{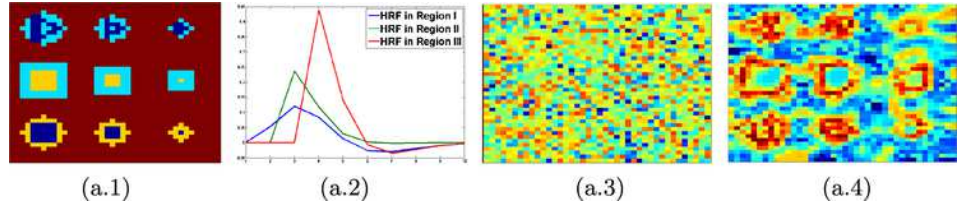}

\caption{The setup of simulation I: \textup{(a.1)} a temporal cut of the
true activation pattern; \textup{(a.2)} the true HRFs with $H_1(t)/8$,
$H_2(t)/4$
and $H_3(t)/2$; \textup{(a.3)} a temporal cut of simulated images;
\textup{(a.4)} Gaussian
smoothing of the simulated image. The ground true image has three
different active regions mixed with each other.} \label{figure5}
\end{figure}

The first simulation compared MASM with the frequency method
developed for a single stimulus in \citet{BaiEtal2009}. This
simulation is similar to the one given in \citet{YueEtal2010}. We
simulated a time series with 200 observations according to model
(\ref{MASMeq1}) at each of all 1600 pixels in a $40\times40$
phantom image, which contains 9 separated areas of
activation. These 9 areas were further
grouped into three different patterns with different shapes mixed
together. See Figure \ref{figure5}(a.1), in which the dark blue, sky
blue and yellow colors represent the active Regions I, II and III, denoted
as $R1$, $R2$ and $R3$, respectively. The nonactive region is
%
%
denoted as $R4$. The stimulus function $X(t)$ was generated
according to a boxcar paradigm consisting of either zero or one,
which was independently generated from a Bernoulli random variable
with the success probability 0.15. We set all HRFs to be zero
outside all activation regions, while within each active region
we convolved the boxcar paradigm $X(t)$ with a standard HRF given by
\begin{eqnarray*}
H_j(t)&=&A_j \biggl(\frac{t}{d_{j1}} \biggr)^{a_{j1}}\exp{
\biggl(-{t-d_{j1}\over
b_{j1}} \biggr)}-c \biggl(\frac{t}{d_{j2}} \biggr)^{a_{j2}}\\
&&{}\times\exp{
\biggl(-{t-d_{j2}\over
b_{j2}} \biggr)}{\mathbf1}\bigl(t\in[j-1, 15]\bigr)
\end{eqnarray*}
with
$(A_1,A_2,A_3)=(1,5,3)$, $c=0.35$, $(a_{11},a_{12})=(6,12)$,
$(a_{21},a_{22})=(4,8)$, $(a_{31},a_{32})=(5,10)$,
$(b_{j1},b_{j2})=(0.9,0.9)$, and
$(d_{j1},d_{j2})=(a_{j1}*b_{j1},a_{j2}*b_{j2})$ for $j=1,2,3$. The
signals in each active region were,
respectively, scaled as
\[
Y_1(t)=(H_1\otimes X) (t)/8,\qquad Y_2(t)=(H_2
\otimes X) (t)/4
\]
and
\[
Y_3(t)=(H_3\otimes X)
(t)/2.
\]
%
The error process $\varepsilon(t, {\mathbf d})$ was generated from an AR(1)
given as $\varepsilon(t,{\mathbf d})=0.3\varepsilon(t-1,{\mathbf d})+\xi
$, where
$\xi$ is a pure Gaussian noise $N(0,\sigma^2)$ with
$\sigma=\sqrt{0.03}$. The simulated BOLD\vadjust{\goodbreak} signals were given by
$Y_j(t,{\mathbf d})+\varepsilon(t,{\mathbf d})$ for $j=1,2,3$. In this
simulation, the
smallest signal-to-noise ratio (SNR) was around 0.5, where SNR is
\mbox{defined} as the mean of the absolute true signals over the standard deviation of
$\varepsilon(t, {\mathbf d})$. We repeated
this simulation 500 times. Figure~\ref{figure5} presents the phantom
image and the simulated image at a single time point with their
related sample curves.

\begin{figure}[t!]

\includegraphics{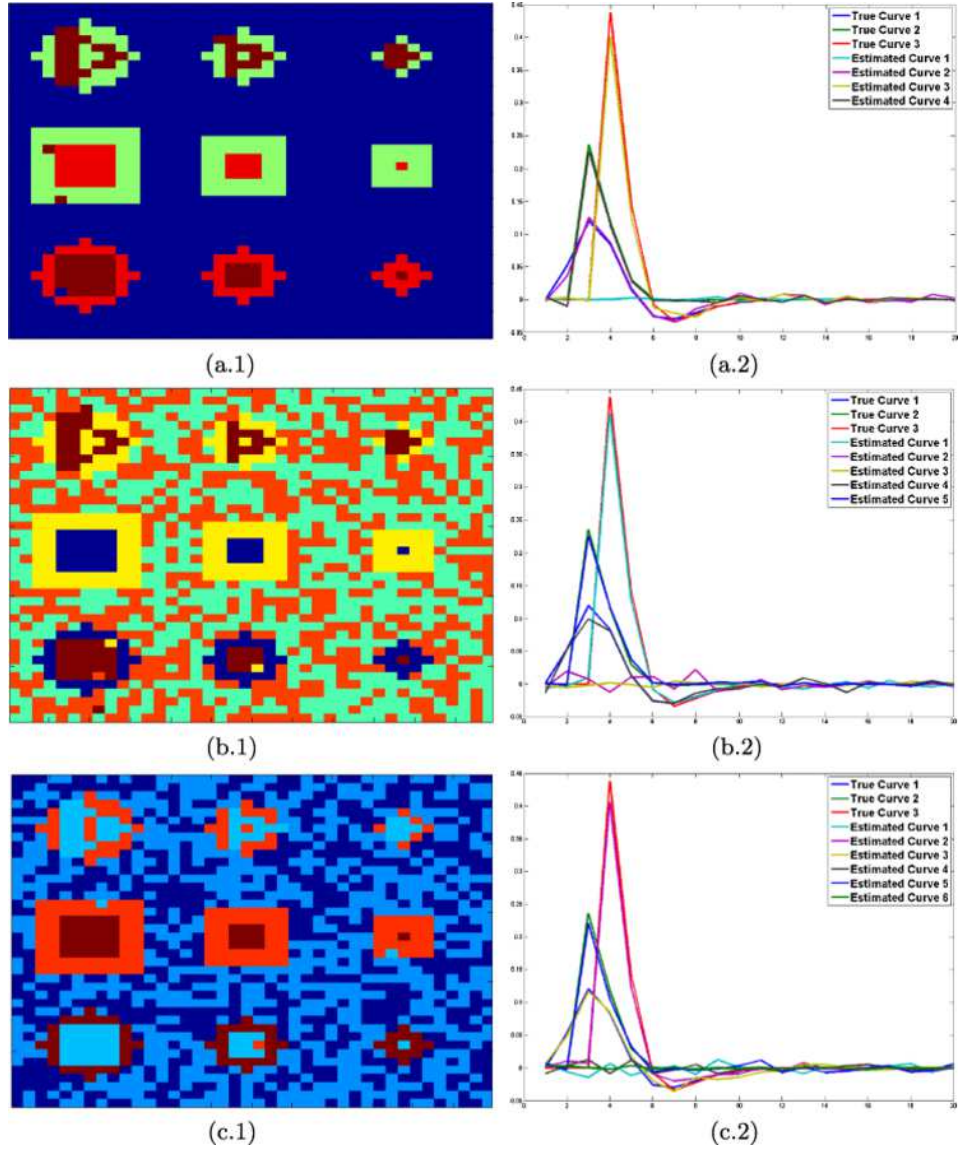}

\caption{The estimated patterns and their mean curves
of HRFs from simulation I. The estimated numbers of clusters may vary
across simulations: \textup{(a.1)} and \textup{(a.2)} number of
clusters${}={}$4; \textup{(b.1)} and \textup{(b.2)} number of
clusters${}={}$5; and \textup{(c.1)} and \textup{(c.2)} number of
clusters${}={}$6. The first column includes the temporal cuts of the
clustering results. The second column includes the averaged estimated
HRFs in their corresponding clustered patterns with the true HRFs,
which are represented by different colors.} \label{figure51}\vspace*{12pt}
\end{figure}

We applied MAE described above to simultaneously estimate HRFs
across all voxels for each simulated data set and then used
the EM-based clustering method to determine the signal pattern and
compute the average
estimator of HRFs in each cluster. Figure \ref{figure51} presents the
clustering patterns with their mean HRFs. Figure \ref{figure51}
reveals several
additional clusters within the nonactive region and their averaged
curves are very close to the zero. This indicates that even
though the number of clusters may vary across simulations, the
activation patterns can be correctly detected. 
The mean estimated HRFs are very close to the ground truth especially
for those
activated regions (see Figure \ref{figure51}).

%

\begin{figure}

\includegraphics{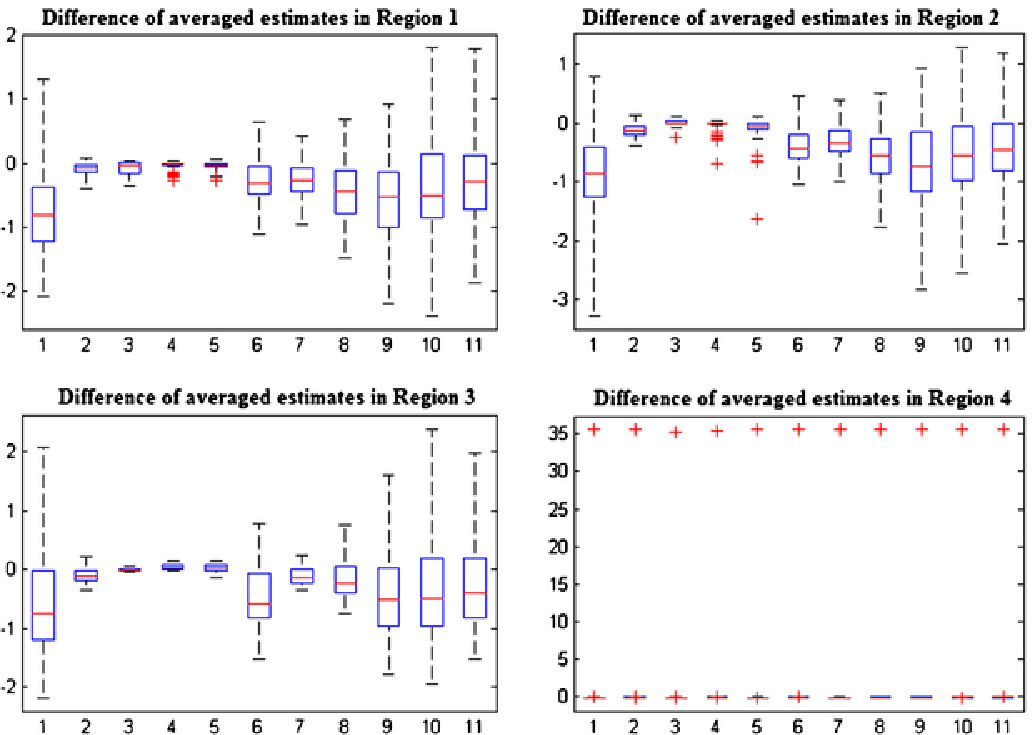}

\caption{The boxplots of AMs (an accuracy measure)
in simulation I: the
differences of the estimated HRFs in the four different regions at
the first 11 time points.} \label{figure6}
\end{figure}

We also applied the woxel-wise frequency domain method of
\citet{BaiEtal2009}, called FMHRF,
to estimate HRFs across all voxels. To compare our method with FMHRF,
we calculated an accuracy measure (AM)
at each of the first 11 time points since these time points represent
the neuronal change at voxel ${\mathbf d}$. The AM is defined as
\[
\operatorname{AM}(t,{\mathbf d})={\sum_{i=1}^{500}\{|x_i(t, {\mathbf d})-H(t,
{\mathbf d})|-|y_i(t, {\mathbf d})-H(t, {\mathbf d})|\}\over
500\cdot\operatorname{Std}(x(t,
{\mathbf d}))}, 
\]
where $x_i(t, {\mathbf d})$ and $y_i(t, {\mathbf d})$ are,
respectively, the
estimated HRFs at time $t$ by using
our method and by
using FMHRF, $\operatorname{Std}(x(t, {\mathbf d}))$ is the standard
deviation of
$\{x_{i}(t, {\mathbf d})\dvtx i=1,\ldots, 500\}$ at time $t$, which is used
to standardize the difference, and $H(t,{\mathbf d})$ is
the corresponding true HRF. A negative value of $\operatorname{AM}(t,{\mathbf
d})$ indicates that the estimated HRFs obtained from our method have
smaller bias compared to FMHRF. Figure \ref{figure6} reveals that
our method
outperforms FMHRF at almost all time points.
In Figure \ref{figure6}, we note an outlier in $R4$, which may be
caused by over-smoothing in some boundary voxels.

%

\begin{figure}[b]

\includegraphics{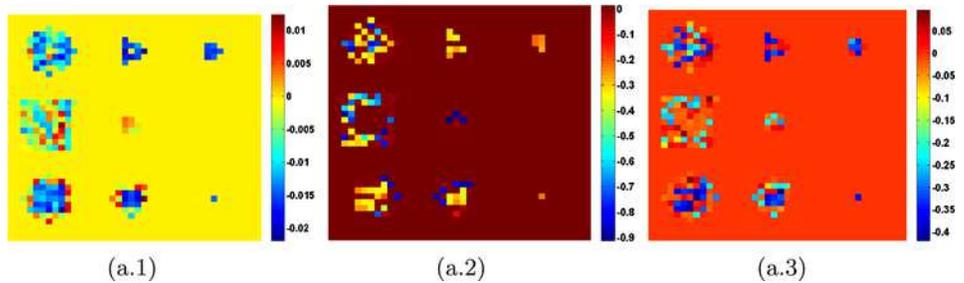}

\caption{The comparison statistics $D_{\mathbf d}$ in simulation I based
on \textup{(a.1)} the estimated height
($H_a$); \textup{(a.2)} the estimated time-to-peak ($T_p$); and \textup{(a.3)} the estimated
width (W) at each active voxel. The color bar denotes the value of
$D_{\mathbf d}$ at each voxel.} \label{figure12}
\end{figure}

We used an isotropic Gaussian kernel with FWHM 5 mm to smooth the
simulated imaging data
and applied FMHRF to the Gaussian smoothed
data. Then, we compared the obtained results with those calculated from
MASM based on the nonsmoothed imaging data.
We compared
MAE and FMHRF by calculating the differences between the estimated
$H_a$, $T_p$ and $W$ and their corresponding true values. Specifically,
for $H_a$,
$T_p$ and $W$, a comparison statistic in the ${\mathbf d}$th voxel is
defined by
\[
D_{\mathbf d}={1\over500}\sum_{i=1}^{500}\bigl(|
\hat{x}_{i, \mathbf d}-x_{0, \mathbf
d}|-|\hat{y}_{i, \mathbf d}-x_{0, \mathbf d}|\bigr),
\]
where $x_{0, \mathbf d}$ represents the true value of $H_a$, $T_p$ or $W$
and $\hat{x}_{i, \mathbf d}$ and $\hat{y}_{i,\mathbf d}$ represent
the estimated $H_a$,\vadjust{\goodbreak}  $T_p$ or $W$ obtained from MASM and FMHRF,
respectively, at
voxel~${\mathbf d}$. A negative value of $D_{\mathbf d}$ indicates that
the estimated
HRFs obtained from MASM are closer to the true value compared to
FMHRF, since standard Gaussian smoothing can blur the BOLD signals
in the boundary voxels of active regions, especially those regions
with a small number of voxels. Figure \ref{figure12}
reveals that MASM outperforms FMHRF in the smallest active regions
and the lowest SNR for all three parameters, especially $H_a$ and
$W$.

%
%


\subsection{Simulation II: Multiple stimuli}\label{sec3.2}

\begin{figure}

\includegraphics{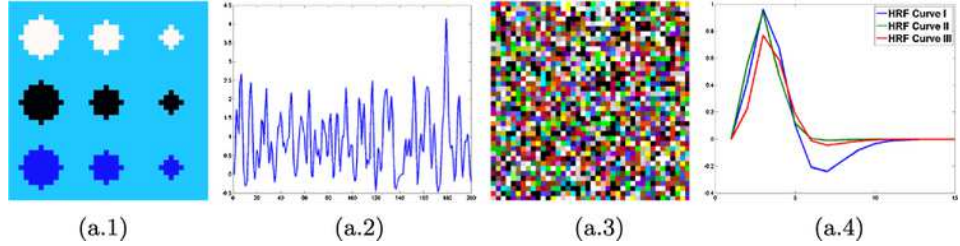}

\caption{The setup of simulation II: \textup{(a.1)} a temporal cut of
the true images; \textup{(a.2)} the true BOLD signals $Y(t)$; \textup{(a.3)} a temporal
cut of the
simulated images; and \textup{(a.4)} the true curves of HRF: $H_1(t)$, $H_2(t)$,
and $H_3(t)$, which are scaled
into three different values representing the different active
regions corresponding to the three stimuli.} \label{figure8}
\end{figure}

The second simulation compared MASM with several state-of-the-art
methods discussed in \citet{LindquistEtal2009}. This fMRI simulation
is similar to the first one except that we consider three stimuli.
We simulated the data with 200 time points (i.e., $T=200$) in a
$40\times40$ phanton image containing 9 regions of
activation-circles with varying radii and a background region with
zeros at each time point. These 9 active regions were also grouped
into three different BOLD patterns with each group consisting of
three circles, which had the same true signal series.
The three true HRFs were defined as 
%
\begin{eqnarray*}
H_j(t)&=&A_j \biggl(\frac{t}{
d_{j1}} \biggr)^{a_{j1}}\exp{ \biggl(-\frac{(t-d_{j1})}{b_{j1}}
\biggr)}\\
&&{}-c \biggl(\frac{t}{
d_{j2}} \biggr)^{a_{j2}}\exp{ \biggl(-\frac{(t-d_{j2})}{ b_{j2}}
\biggr)}{\mathbf
1}\bigl(t\in[0, 15]\bigr)
\end{eqnarray*}
with $(A_1,A_2,A_3)=(1,5,3)$,
$c=0.35$, $(a_{11},a_{12})=(6,12)$, $(a_{21},a_{22})=(4,8)$,
$(a_{31},a_{32})=(5,10)$, $(b_{j1},b_{j2})=(0.9,0.9)$, and
$(d_{j1},d_{j2})=(a_{j1}*b_{j1},a_{j2}*b_{j2})$ for $j=1,2,3$. The
boxcars (e.g., the stimulus sequence) consisting of either zero or
one were randomly generated by a Bernoulli trial independently with
the successful rate 0.15 for $j=1,2,3$. The true BOLD signals were
calculated according to $Y(t)=\sum_{j=1}^3(H_j\otimes X_j)(t)$. The
signals in the three activation-circle groups were then scaled to
be $Y_1(t)=Y(t)/6$, $Y_2(t)=Y(t)/4$ and $Y_3(t)=Y(t)/2$,
respectively. The noise terms $\varepsilon(t, {\mathbf d})$ were generated
from a
Gaussian distribution $N(0,\sigma^2)$ with $\sigma=0.2$. Finally, the
simulated BOLD signals were set as $Y(t, {\mathbf d})+\varepsilon(t,
{\mathbf d})$ for $j=1,2,3$.
In this simulation, the smallest SNR was around~0.6. The background
and the simulated images with their related curves at some time
points are given in Figure \ref{figure8}.

%

\begin{figure}

\includegraphics{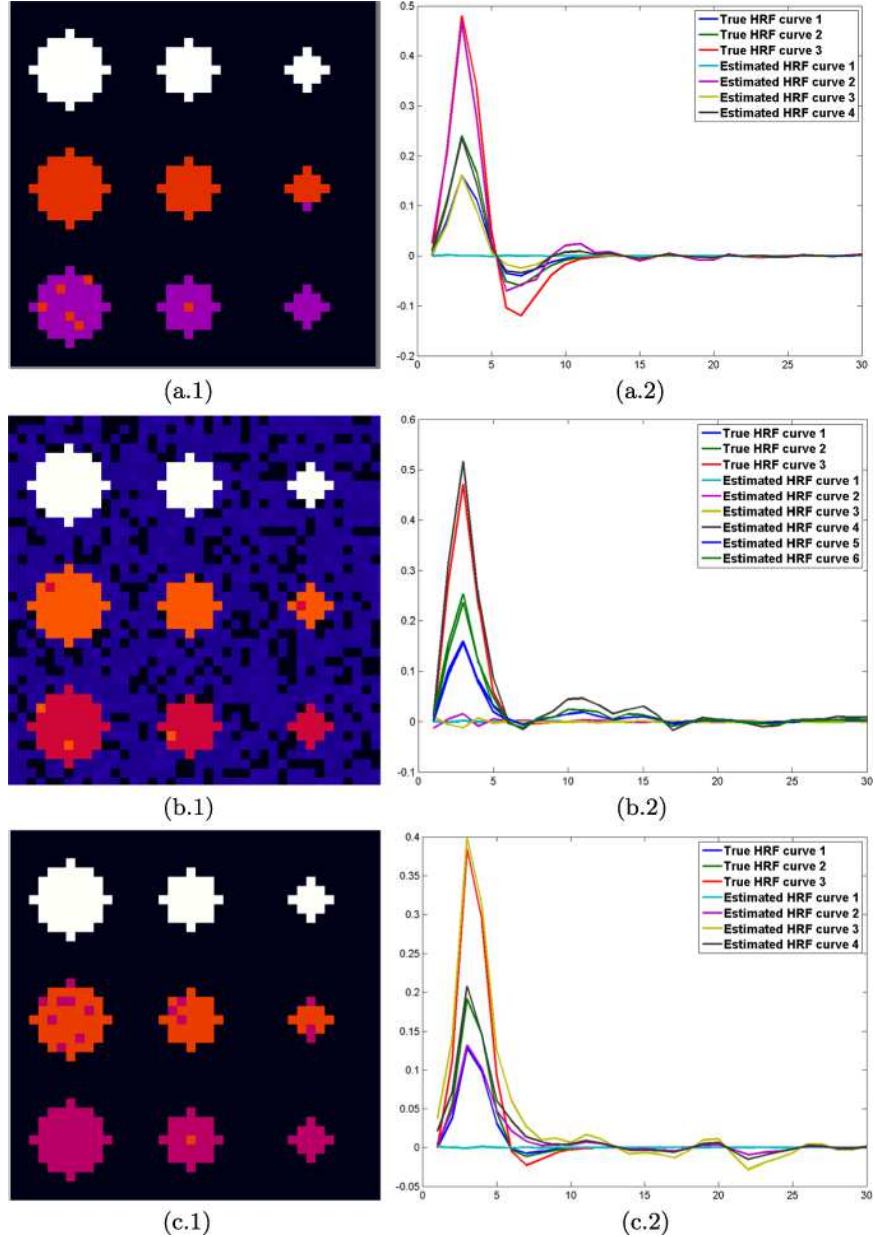}

\caption{The estimated patterns and the mean curves
of HRFs in their patterns for simulation II. The estimated patterns and
their mean curves for
the first stimulus sequence \textup{(a.1)} and \textup{(a.2)}; the second
stimulus sequence \textup{(b.1)} and \textup{(b.2)}; and the third stimulus sequence
\textup{(c.1)} and \textup{(c.2)}. The column \textup{(a.1)}, \textup{(b.1)}, \textup{(c.1)} includes the temporal cuts of the
clustering results. The column \textup{(a.2)}, \textup{(b.2)}, \textup{(c.2)} includes the averaged
estimated HRFs in their corresponding clustered patterns with the
true HRFs, which are represented by using different colors. The
numbers of the clusters also vary across simulations for each
stimulus sequence.} \label{figure81}
\end{figure}

We applied our MAE to simultaneously estimate all HRFs across all
voxels in each of 500 simulated data sets. Then we clustered the
estimated HRFs by using the EM
algorithm 
and calculated the mean curves of all patterns. See
Figure \ref{figure81}, in which the estimated
HRFs corresponding to
the three stimulus sequences are presented. 

%

%
\begin{figure}

\includegraphics{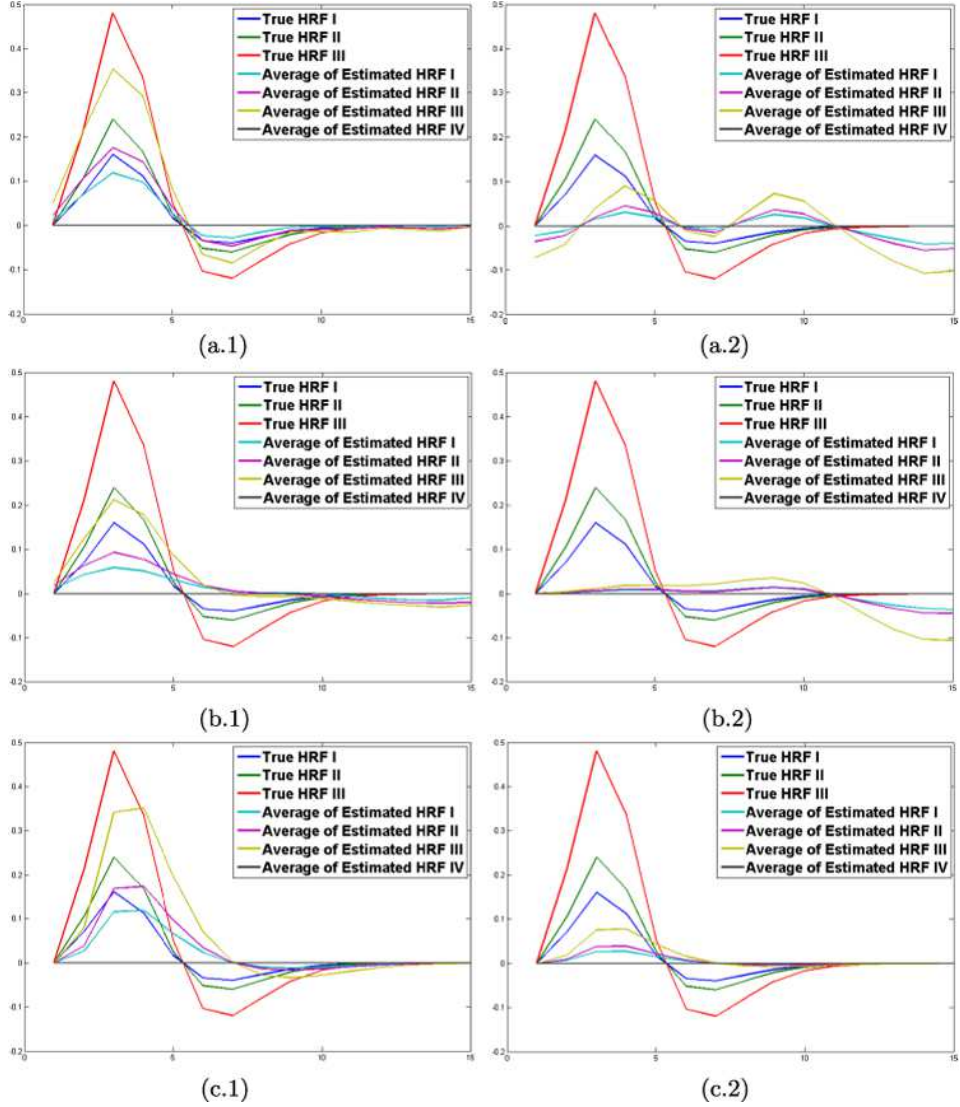}

\caption{The mean HRFs in each region from the first stimulus
sequence of one sample in simulation~II, estimated from sFIR, IL and
GAM based on the smoothed or nonsmoothed data: \textup{(a.1)}, \textup{(b.1)}, \textup{(c.1)} the
averaged estimated HRFs from raw data; and \textup{(a.2)}, \textup{(b.2)}, \textup{(b.2)} the averaged
estimated HRFs from smoothed data. \textup{(a.1)}, \textup{(a.2)} mean HRFs estimated
from sFIR; \textup{(b.1)}, \textup{(b.2)} mean HRFs estimated from IL; and \textup{(c.1)}, \textup{(c.2)} mean
HRFs estimated from GAM.} \label{figure131}
\end{figure}

We considered three state-of-the-art methods discussed in
\citet{LindquistEtal2009} including the following: (i) SPMs
canonical HRF (denoted
as GAM), which is a parametric approach by assuming the HRF is a
mixture of Gamma functions; (ii) the finite impulse response (FIR)
basis set, named as
the semi-parametric smooth FIR model (sFIR), which assumes that HRF can
be estimated by a linear combination of some basis functions; and (iii)
the inverse
logit model (IL), which considers the HRF as a linear combination of
some inverse logit functions. As a demonstration of the mean curves in each
region estimated from these methods, we only display the results
from one stimulus in one sample in Figure \ref{figure131}, from which
we can find the estimated HRFs from either smoothed or nonsmoothed
data are over-smoothed even though they have a similar trending pattern
as the true HRFs.

\begin{figure}

\includegraphics{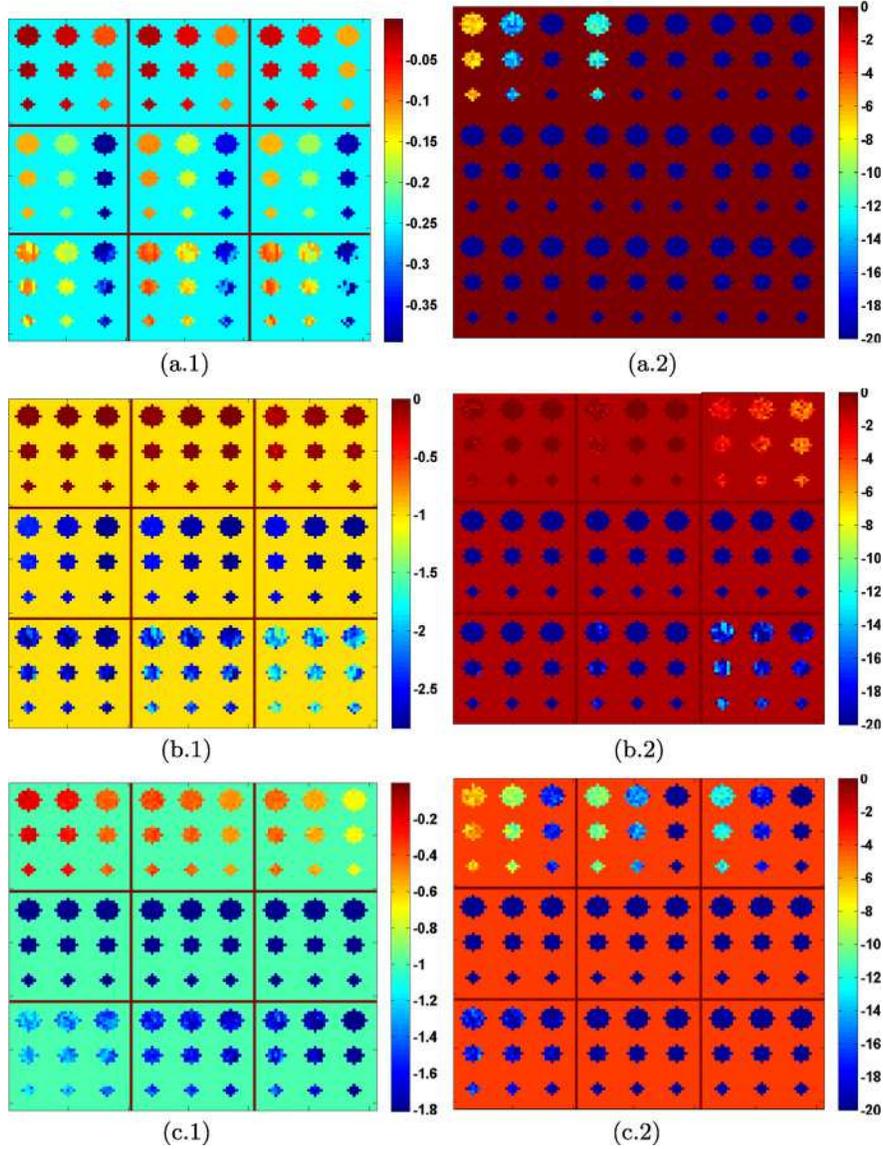}

\caption{The values of $D_{\mathbf d}$ and paired $t$-test statistics in
simulation II: the values of $D_{\mathbf d}$ for the estimated \textup
{(a.1)} height
($H_a$); \textup{(b.1)} time-to-peak ($T_p$); and \textup{(c.1)} width
($W$); and paired
$t$-test statistics for the estimated \textup{(a.2)} height ($H_a$);
\textup{(b.2)}
time-to-peak ($T_p$); and \textup{(c.2)} width ($W$)
at each active voxel for the three stimulus sequences. For panels
\textup{(a.1)}, \textup{(b.1)} and \textup{(c.1)}, the 1st, 2nd and 3rd
rows are the average
values of $D_{\mathbf d}$ between MASM and sFIR, between MASM and GAM, and
between MASM and IL, respectively.
For panels \textup{(a.2)}, \textup{(b.2)} and \textup{(c.2)}, the 1st,
2nd and 3rd rows are paired
$t$-test statistics between MASM and sFIR, between MASM and GAM, and
between MASM and IL, respectively.
In each panel, the 1st, 2nd and 3rd columns come from the 1st, 2nd and
3rd stimulus sequences, respectively. The paired $t$-statistics are
truncated at $-20$.}
\label{table3}
\end{figure}

These over-smoothed results also can be reflected in the following
statistics. Based on the estimated HRF, we computed $H_a$, $T_p$ and
$W$ as the potential measure of response magnitude, latency and
duration of neuronal activity, respectively. We compared our method
with sFIR, IL and GAM based on the differences between the estimated
statistics $H_a$, $T_p$ and $W$ and the true ones. We also calculated the
evaluation statistics $D_{\mathbf d}$ for the ${\mathbf d}${th} voxel.
Figure \ref{table3} indicates that our method can provide more
accurate estimates of the HRF statistics, compared with all others,
especially GAM and IL. Moreover, most values of $D_{\mathbf d}$ are
negative and
statistically significant at the 0.05 significance level.
Also, the average of the differences between MASM and sFIR is
small
in the estimation of $H_a$, $T_p$ and $W$.

%



%
\begin{figure}

\includegraphics{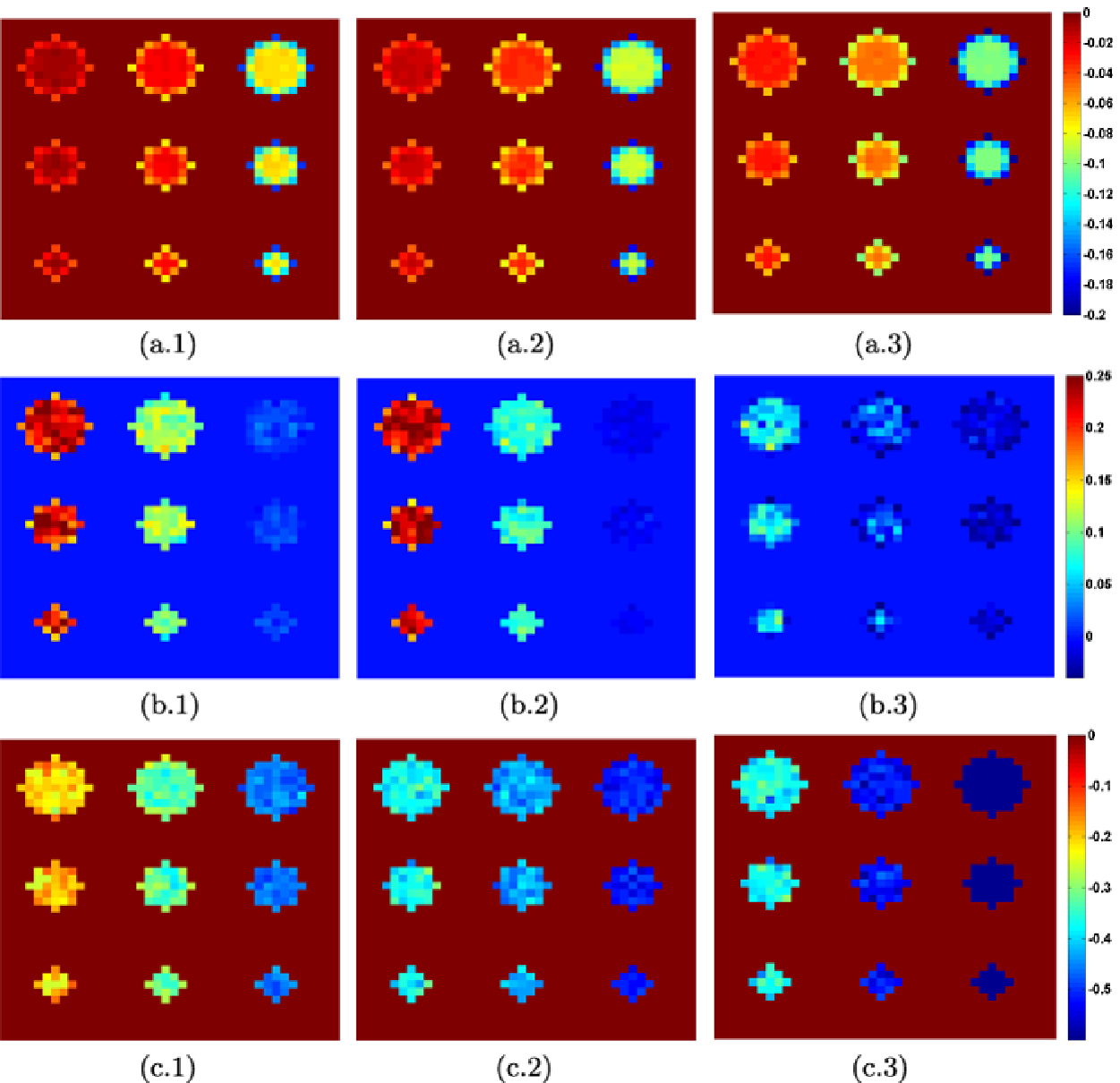}

\caption{The comparison statistics $D_{\mathbf d}$ in simulation II with
sFIR based on \textup{(a.1)}--\textup{(a.3)} estimated height ($H_a$); \textup{(b.1)}--\textup{(b.3)}
estimated time-to-peak ($T_p$); and \textup{(c.1)}--\textup{(c.3)} estimated
width ($W$) at each active voxel for the three stimulus sequences. The color
bar denotes the value of $D_{\mathbf d}$ for the ${\mathbf d}$th voxel.}
\label{figure13}
\end{figure}

\begin{figure}

\includegraphics{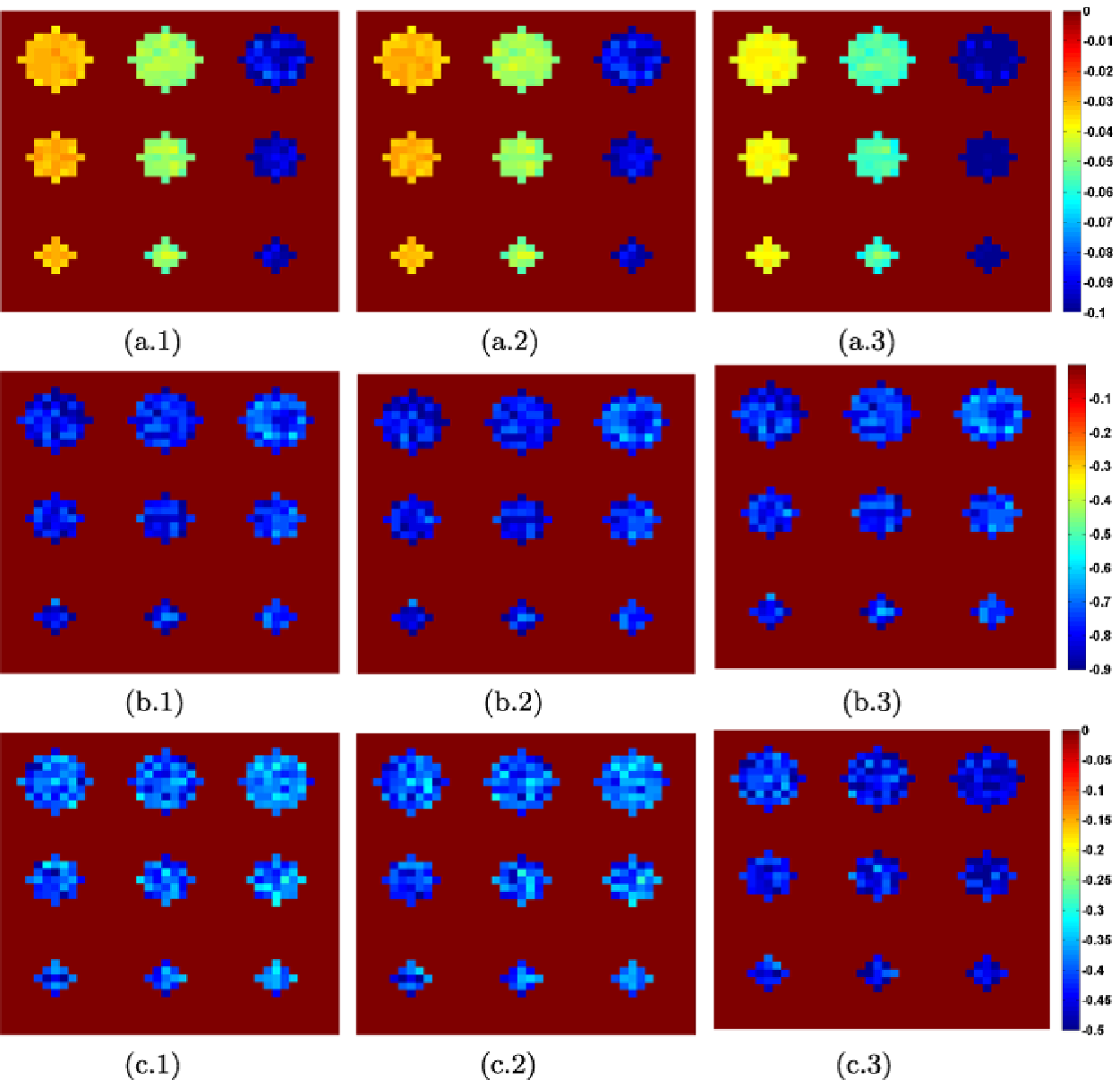}

\caption{The comparison statistics $D_{\mathbf d}$ in simulation II with
IL based on \textup{(a.1)}--\textup{(a.3)} estimated height ($H_a$); \textup{(b.1)}--\textup{(b.3)} estimated
time-to-peak ($T_p$); and \textup{(c.1)}--\textup{(c.3)} estimated
width ($W$) at each active voxel for the three stimulus sequences.}
\label{figure15}
\end{figure}

\begin{figure}

\includegraphics{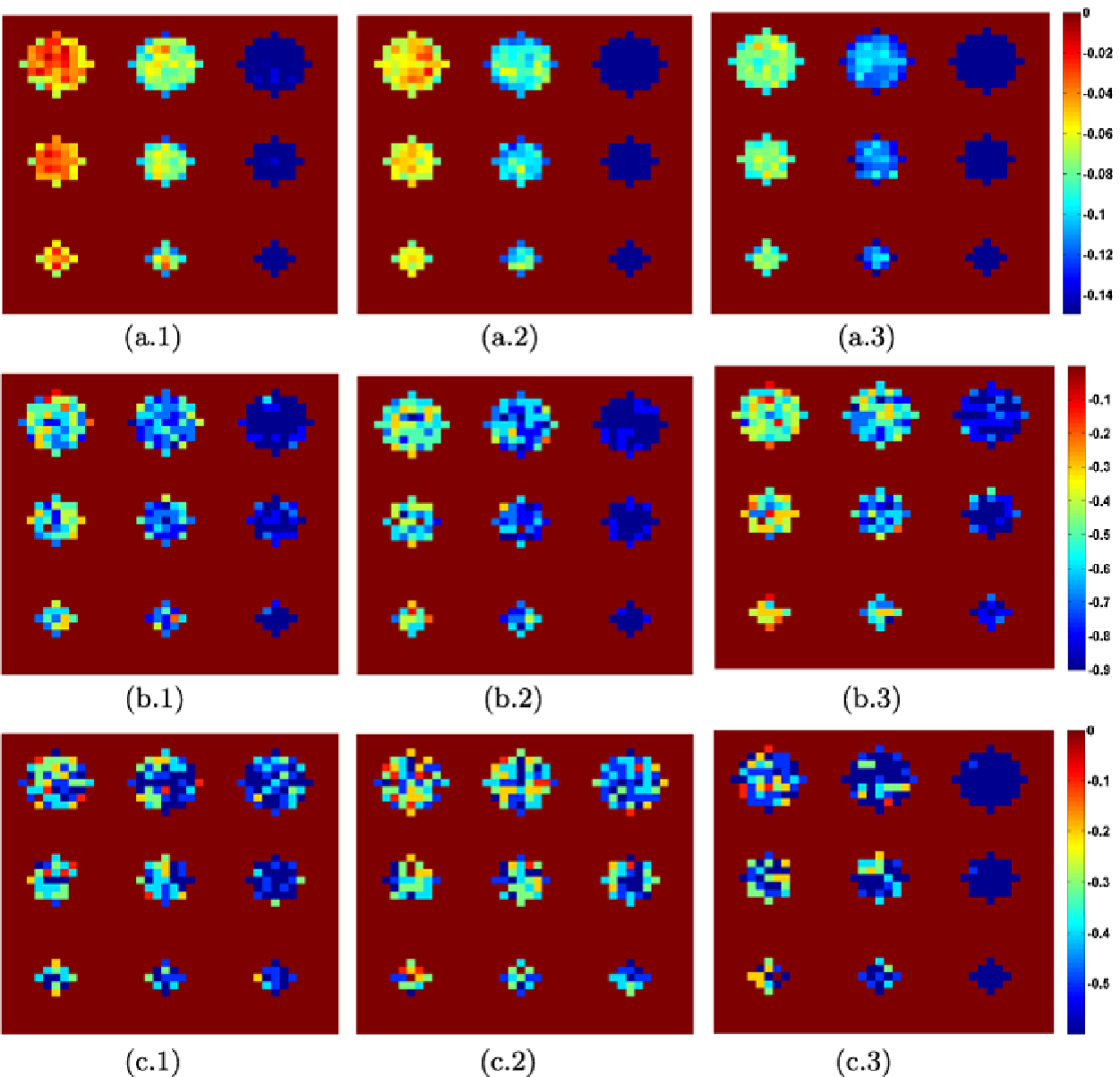}

\caption{The comparison statistics $D_{\mathbf d}$ in simulation II with
GAM based on \textup{(a.1)}--\textup{(a.3)} estimated height ($H_a$); \textup{(b.1)}--\textup{(b.3)}
estimated time-to-peak ($T_p$); and \textup{(c.1)}--\textup{(c.3)} estimated
width ($W$) at each active voxel for the three stimulus sequences.}
\label{figure14}
\end{figure}

We also applied the Gaussian smoothing with FWHM equal 5 mm to the
simulated imaging data before running sFIR, IL and GAM and then we
compared them to MASM based on unsmoothed data.
Figure \ref{figure13} reveals that
MASM outperforms sFIR in the estimation of $H_a$ and $W$, but not
$T_p$. This is consistent with the comparison in Figure \ref{table3}.
Figures \ref{figure15} and \ref{figure14} reveal that the
differences $D_{\mathbf d}$ for all three parameters of interest
are negative in almost all voxels of the activation regions.
This indicates that MASM outperforms sFIR, IL and GAM,
even after applying the Gaussian smoothing.

%

%

Finally, we computed the computation times for sFIR, IL, GAM and MASM,
which are shown in
Table \ref{table2}.
Although MASM uses the information from neighboring voxels, its
computation time slightly increases compared to GAM. As expected, the
computation time of MASM is longer than that of GAM and sFIR, but
shorter than that of IL.

\section{Data analysis}\label{sec4}

To examine the finite sample performance of our MASM on real fMRI data,
we used a
fMRI data set collected from a study designed to test the hypothesis that
implicit retrieval of conceptual and perceptual associations is
differentially linked with medial temporal lobes (MTL). In this
study, 19 subjects completed an associative version of a speeded
classification task, in which they decided which of two objects was
more likely to be found inside a house.
We first chose some regions of interest in the implicit test fMRI data
from a randomly selected subject to examine the estimation accuracy of MASM,
and then we computed the images of height,
time-to-peak and width from all subjects
to compare the group-wise differences between MASM and three other
competing methods.

The stimuli were 180 line drawings of familiar objects taken from
the Microsoft online clip art database at the website
\href{http://www.clipart.com/en/}{www.clipart.com}. Each object was
filled in with a single, plausible
color using Adobe Photoshop. Objects were pilot-tested for
consistency in response to the associative classification task (an
inside/outside judgment). Critical trials consisted of two objects
presented side by side. The implicit test consisted of the 42
studied trials, 14 of which were presented as intact pairs (objects
studied together), 14 were recombined (each object studied but not
together) and 14 were recolored versions of otherwise intact pairs.
Each new color was a plausible real-world color for any given
object. The implicit test also included 14 new, unstudied pairs as
well as 26 null trials. So there are in total 4 sequences of the
stimuli. Finally, the null trials were used to assess baseline
activation levels.

%
%
\begin{table}
\tablewidth=240pt
\caption{Comparisons of average computing times (in
seconds) in the same computer but with the different programming
environments. sFIR, IL and GAM are written in Matlab and MASM in the
computer language C}
\label{table2}
\begin{tabular*}{\tablewidth}{@{\extracolsep{\fill}}l ccc c@{}}
\hline
& \textbf{sFIR} & \textbf{IL} & \textbf{GAM} & \textbf{MASM} \\
\hline
One stimulus &1.47 &2934.6&\hphantom{0}5.31&\hphantom{0}67.33\\
Three stimuli& 3.04&9927.3 & 13.74& 219.0\hphantom{0}\\
\hline
\end{tabular*}
\end{table}

\subsection{Data acquisition}\label{sec4.1}
Whole-brain gradient-echo, echo-planar images\break were collected
(forty-six 3 mm slices, TR${}={}$3 s, TE${}={}$23 ms) using a 3T Siemens Allegra
scanner while the participants performed the cognitive task. Slices
were oriented along the long axis of the hippocampus with a
resolution of 3.125~mm${}\times{}$3.125 mm${}\times{}$3 mm. High-resolution T1-weighted
(MP-RAGE) structural images were collected for anatomic
visualization. Stimuli were back-projected onto a screen and viewed
in a mirror mounted above the participant's head. Responses were
recorded using an MR-compatible response box. Head motion was
restricted using a pillow and foam inserts.

\subsection{Analysis results}\label{sec4.2}
We used SPM [see \citet{FristonEtal2009}] to preprocess the fMRI data,
including the realignment, timing slicing, segmentation, coregister,
normalization and spatial smoothing.
To de-trend the data, we implemented a
global signal regression method which can enhance the quality of the
data and remove the spontaneous fluctuations common to the whole
brain [see \citet{MurphyEtal2009}].
Then in the first analysis, we used a canonical
HRF model with time and dispersion derivatives to estimate the HRFs
corresponding with the four sequences of the stimulus events. In the
2nd level estimation of SPM, $F$-statistic maps were computed to
detect the activation/deactivation regions triggered by the four
stimuli and then we set a threshold with the raw $p$ value less
than 0.01 and the extension $K=20$ to find the significant
regions of interest (ROIs).
To evaluate the performance of MASM,
we randomly selected a significant ROI detected by SPM for each
stimulus type and calculated HRFs and their associated statistics by using
all four HRF estimation methods based on fMRI data in
each ROI.



%

\begin{figure}

\includegraphics{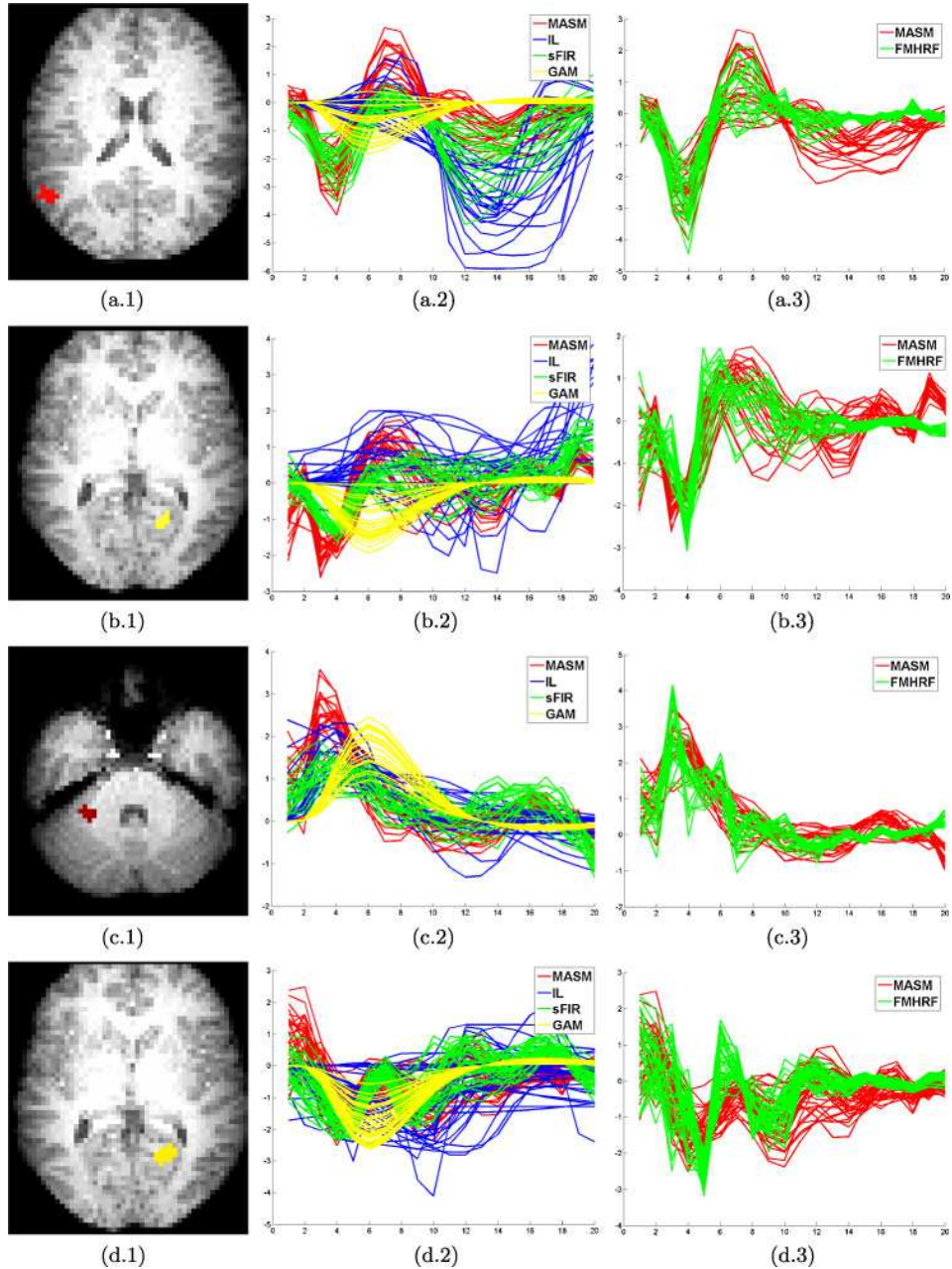}

\caption{The fMRI data analysis results: \textup{(a.1)}, \textup{(b.1)}, \textup{(c.1)}, \textup{(d.1)} the
selected slices
of the $F$-statistic maps with significant ROIs for the
1st--4th stimulus sequences from the top to the
bottom, in which red, yellow and brown colors represent
the selected significant ROIs; \textup{(a.2)}, \textup{(b.2)}, \textup{(c.2)}, \textup{(d.2)}
estimated HRFs in the significant ROIs corresponding to each stimulus
from MASM (red), IL (blue), sFIR (green) and GAM (yellow); \textup{(a.3)}, \textup{(b.3)},
\textup{(c.3)}, \textup{(d.3)}
estimated HRFs from MASM (red) and FMHRF (green) in the significant
ROIs.} \label{figure11}
\end{figure}

We presented the estimated HRFs from all four HRF estimation methods in
Figure \ref{figure11} and compared their shapes.
Figure \ref{figure11} reveals that the shape of estimated HRFs from
GAM, sFIR and MASM is consistent with the pattern of the selected
%
%
activation and deactivation ROIs. However, as shown in Figure \ref
{figure11}(b.2) and (d.2), it seems that IL does not work well in the
deactivation ROIs, since there is a large variation of the estimated
HRFs from IL.
The HRF parameters including $H_a$, $T_p$ and $W$ obtained from MASM
and sFIR differ significantly from those obtained from GAM, since GAM
as a
parametric model may not be flexible enough to capture the shape of
true HRFs. This result is also consistent with our simulation
results in Figure \ref{table3}, that is, the differences between
sFIR and MASM are much smaller than those between GAM and MASM and
between IL and MASM. On the other hand, sFIR has larger variability
in the tail of estimated HRFs and smaller height compared to MASM. It may
indicate that MASM provides more accurate estimation of HRF and its
associated parameters compared with
GAM, IL and sFIR.

We compared the results of MASM with those of FMHRF, which are
presented in Figure \ref{figure11}. Figure \ref{figure11}
shows that the estimated HRFs from MASM and FMHRF have similar
profiles. However, compared with FMHRF,
the estimated HRFs from MASM look
smoother and can capture more
dynamic changes at their tails. This may be due to the fact that FMHRF
only uses fMRI data at each voxel, whereas
MASM adaptively incorporates fMRI data from the neighboring information
of each voxel. If we could treat the estimated HRFs from
sFIR as the ground truth, the estimated HRFs from MASM are closer to
those from sFIR than
those from FMHRF.

\begin{figure}

\includegraphics{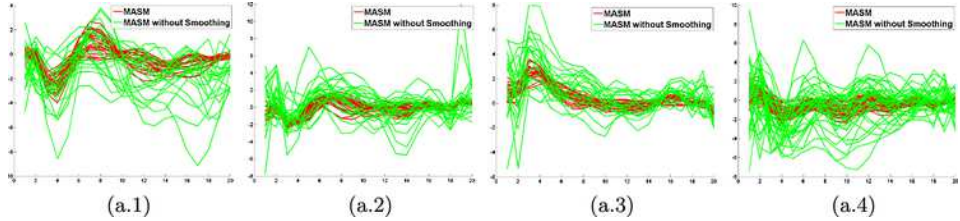}

\caption{The fMRI real data analysis results:
\textup{(a.1)}--\textup{(a.4)} estimated HRFs from MASM based on the smoothed fMRI data
(red) and based on the ``raw'' fMRI data (green) in each
ROI.} \label{figure101}
\end{figure}

Finally, we applied MASM to the ``raw'' fMRI data without using the
Gaussian smoothing step in the preprocessing pipeline. We used the same
set of parameters in MASM to estimate HRFs and compared them with those
from MASM based on the smoothed fMRI data. See Figure \ref{figure101}
for detailed comparisons. Figure \ref{figure101} reveals that
although the estimated HRFs from the raw and smoothed fMRI data have
similar shape, their amplitudes based on the raw fMRI data are larger
than those based on the smoothed fMRI data since the use of Gaussian
smoothing can reduce the amplitudes of estimated HRFs.

%

%
\begin{figure}[b]

\includegraphics{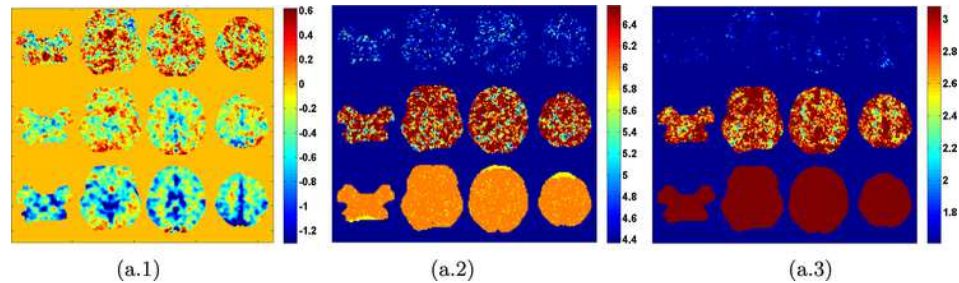}

\caption{The fMRI real data analysis results: the mean images of the estimated
\textup{(a.1)} height ($H_a$); \textup{(a.2)} time-to-peak ($T_p$); and \textup{(a.3)} width
($W$) at some selected slices. The first row is from MASM; the second
row is from sFIR; and the third row is from GAM.} \label{figure102}
\end{figure}

%
\begin{figure}

\includegraphics{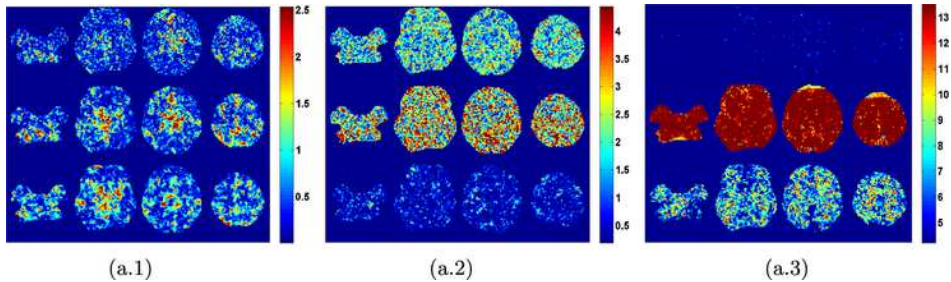}

\caption{The fMRI real data analysis results:
the $-\log_{10}(p)$ images for testing the differences of the
estimated \textup{(a.1)} height ($H_a$); \textup{(a.2)} time-to-peak ($T_p$); and \textup{(a.3)}
width ($W$) across different methods at some selected slices. The first
row is from the differences between MASM and sFIR; the second row is
from the differences between MASM and GAM; the third row is from the
differences between sFIR and GAM.}
\label{figure103}
\end{figure}

We also calculated the three estimated HRF parameters from MASM, sFIR
and GAM for all subjects and then compared them across different
methods. Note that we omitted IL here due to its bad performance in
those deactivated ROIs. For the sake of space, we only included the
estimated HRFs from the first stimulus sequence from all subjects. See
additional results from other stimulus sequences in Part E of the
supplementary material [\citet{Jiaetal}]. Figure \ref{figure102}
shows the
mean images of $H_a$, $T_p$ and $W$ calculated from different methods
in four selected slices. Figure \ref{figure103} displays the
$-\log_{10}(p)$ maps for statistically comparing MASM with sFIR, MASM
with GAM, and sFIR with GAM by using the paired $t$-test. Figure
\ref{figure102}(a.1) reveals that although the heights from MASM are
larger than those from sFIR and GAM, their values are closer to those
from sFIR than those from GAM for most voxels. This is consistent with
the results in Figure \ref{figure103}(a.1). In contrast, Figure
\ref{figure102} reveals that the time-to-peaks and widths from MASM are
smaller than those from sFIR and GAM. For the width, as shown in Figure
\ref{figure103}(a.3), the difference between MASM and sFIR is smaller
than those between MASM and GAM and between sFIR and GAM. This is also
consistent with the simulation studies (see Figure \ref{table3}). As
shown in Figures \ref{figure102} and \ref{figure103}, in many voxels,
the estimated HRFs from MASM have short delay and quick decay, but
large amplitude, whereas those from sFIR have long delay and slow
decay, but small amplitude. It may indicate that MASM outperforms sFIR
in this fMRI data set.



\section{Conclusion and discussion}\label{sec5}

This paper has developed a multiscale adaptive smoothing model
to spatially and simultaneously estimate HRFs
for the BOLD signals
across all
voxels. 
MASM is a nonparametric estimation procedure,
which is shown to be self-calibrating and accurate when compared to
other approaches in the time domain, including the standard methods in
SPM. Also, compared with the method in
\citet{BaiEtal2009} and those in \citet{LindquistEtal2009}, our
approach can provide more accurate and precise estimates of HRFs by
involving the local spatial and frequency information, as shown in
the two simulations and the real data analysis. Moreover, MASM does
not assume any parametrical form and is useful for justifying the
parametrical models for HRF.

Many issues still merit further research.
The first issue is to deal with weight computation and bandwidth
selection in MASM. Although there are
several weight computation and bandwidth selection procedures in the
fMRI literature, their computational burden can be either intractable
in practice or are developed for different purposes.
For instance, \citet{FrimanEtal2003} developed a constrained
canonical correlation analysis (CCA) to calculate the
weight information between any two curves
in the temporal domain.
Moreover,
\citet{WorsleyEtal1996} proposed an adaptive bandwidth selection
method to perform spatial smoothing for the random field theory.

The second issue is to select the optimal bandwidth in frequency (or
temporal) and spatial domains. One strategy is to separately determine
the optimal bandwidth in each domain and then independently apply them
to fMRI data.
In this case, one can apply the existing methods to select the optimal
bandwidth in either frequency/temporal or spatial domain [\citet
{Lepski1990}, \citet{Lepski1997}, \citet{Donoho1007}]. The other strategy
is to simultaneously select the optimal bandwidth in both frequency (or
temporal) and spatial domains.
In MASM, we use a two-stage strategy consisting of an initial frequency
smoothing step with an initial bandwidth $r_0 = 5/T$ and
a simultaneous smoothing step of expanding the spatial neighborhood
exponentially and the frequency neighborhood linearly. We design such strategy
to balance between estimation accuracy
and computational efficiency for the ultra-high dimensional fMRI data.
Although we have tested such a strategy in both simulation studies and
real fMRI data,
it is unclear whether or not the selected bandwidth is theoretically
optimal, which is a topic of our ongoing research.

The third issue is to develop a unified fMRI pipeline to perform fMRI
data analysis. Such a fMRI pipeline may consist of five key tools,
including MASM for estimating HRFs,
a functional linear model for modeling HRFs across subjects, a testing
procedure for detecting activation sets, a clustering model
for grouping different voxels in ROIs and
a network model for integrating different ROIs into structural and
functional brain hubs.
The other four key tools are topics of our ongoing research. We will
present them elsewhere.



\begin{supplement}
\stitle{Multiscale adaptive smoothing models for the
hemodynamic response function in fMRI}
\slink[doi]{10.1214/12-AOAS609SUPP} 
\sdatatype{.pdf}
\sfilename{aoas609\_supp.pdf}
\sdescription{This document consists of three parts: Part A is the
computation procedure of the test statistics $W^{(l)}(d;h_l,r_l)$; Part
B is the algorithm of EM-based clustering; Part C includes additional
results under different parameter combinations.
Part D are the acronym and notation tables; Part E includes the
additional results from group-wise data analysis.}
\end{supplement}

%

\printaddresses

\end{document}